\documentclass[journal,comsoc]{IEEEtran}
\usepackage[T1]{fontenc}% optional T1 font encoding

  \usepackage[pdftex]{graphicx}

\ifCLASSINFOpdf
  \usepackage[pdftex]{graphicx}
\else
\fi
\usepackage{amsmath}
\usepackage[cmintegrals]{newtxmath}
\usepackage{amsmath}
\usepackage{amsfonts,amssymb}
\usepackage{multirow}

\def\bF{{\mathbb{F}}}

\renewcommand{\P}[1]{{\mathbb{P}}\left(#1\right)}

\newtheorem{Theorem}{Theorem}

\begin{document}
\title{Physical Layer Security for RF Satellite Channels in the Finite-length Regime}
%\title{Design of a Secure Physical Layer for RF Satellite Channels in the Finite-length Regime}
%Information-theoretic Physical Layer Security for Satellite Channels}

\author{\'{A}ngeles V\'{a}zquez-Castro,~\IEEEmembership{Senior Member,~IEEE,}
        and~Masahito~Hayashi,~\IEEEmembership{Fellow,~IEEE}% <-this % stops a space
\thanks{
\'{A}ngeles Vazquez-Castro
is with the Department of Telecommunications and Systems Engineering,
Autonomous University of Barcelona,
Barcelona, Spain
e-mail: angeles.vazquez@uab.es.}% <-this % stops a space
\thanks{Masahito Hayashi is with
Graduate School of Mathematics, Nagoya University, Nagoya, Japan and
Centre for Quantum Technologies, National University of Singapore, Singapore.
e-mail: masahito@math.nagoya-u.ac.jp}% <-this % stops a space
\thanks{Manuscript submitted 30th June 2017; revised xxx, 2017.}}


\markboth{Journal of \LaTeX\ Class Files,~Vol.~14, No.~8, August~2015}%
{Shell \MakeLowercase{\textit{et al.}}: Bare Demo of IEEEtran.cls for IEEE Communications Society Journals}



\maketitle

\begin{abstract}
Secure communications is becoming increasingly relevant in the development of space technology. 
Well established cryptographic technology is already in place and is expected to continue to be so. On the other hand, information theoretical security emerges as a post-quantum versatile candidate to complement overall security strength. In order to prove such potential, performance analysis methods are needed that consider realistic legitimate and eavesdropper system assumptions and non-asymptotic coding lengths.

In this paper we propose the design of secure radio frequency (RF) satellite links with realistic system assumptions. Our contribution is three-fold. First, we propose a wiretap channel model for the finite-length regime. The model includes an stochastic wiretap encoding method using existing practical linear error correcting codes and hash codes. Secrecy is provided with privacy amplification, for which the finite-length secrecy metric is given that upper bounds semantic secrecy. 
Second, we derive a novel RF (broadcast) satellite wiretap channel model that parameterizes the stochastic degraded channel around the legitimate channel, a necessary condition to enable secure communication. Finally, we show the design of a secure satellite physical layer and finite-length performance evaluation. In doing so, we define as \emph{sacrifice rate} the fixed fraction of the overall coding rate budget for reliability that needs to be allocated to secrecy. Our methodology does not make use of channel side information of the eavesdropper, only assumes worst case system assumptions. We illustrate our proposed design method with numerical results using practical error correcting codes in current standards of satellite communication.
\end{abstract}

% Note that keywords are not normally used for peer review papers.

 \ifCLASSOPTIONpeerreview
 
 \begin{center} \bfseries EDICS Category: 3-BBND \end{center}
 \fi
%here the extra information on the cover for peer review papers.

\IEEEpeerreviewmaketitle

\section{Introduction}

\subsection{Motivation}

\IEEEPARstart{I}{nterest} in ensuring a desired level of security in communication over space links has been steadily increasing in the last years. Widely used protocols of the Consultative Committee for Space Data Systems (CCSDS) Packet TM/TC protocol family already includes basic authentication, encryption and authenticated encryption \cite{CCSDS_Cryp1}\cite{CCSDS_Cryp2}\cite{CCSDS_Cryp3}. These solutions are mainly available at the link layer and mostly targeting protection of the telemetry and/or telecommand links for current space missions security architectures. These usually have simple topologies consisting of a single spacecraft and a ground segment connected to multiple ground stations, a command and control centre and a Payload Data Ground Segment (PDGS)  \cite{Ignacio2010}. End-to-end techniques can therefore extend the data protection to the entire flow of user data. 
On the other hand, advanced network architectures are currently under study and will possibly include networked or federated satellites. This increase in complexity requires that the security architecture and corresponding security protocols must evolve accordingly. Further, technological advances are also under development at physical and system levels. For example, upcoming software-defined radio and networking paradigms as well as adaptive physical layer mechanisms will surely modify the operational assumptions of security mechanisms. Finally, security threads and types of attacks are ever evolving calling for stronger and complementary security solutions that could be jointly operative.

Within such context, physical layer security based on information theoretical security and cryptographic methods emerges as a post-quantum versatile candidate to complement overall system security strength. In order to prove such potential, performance analysis methods are needed that consider realistic legitimate and eavesdropper system assumptions. In the following, we review the state of the art in the area and outline our contributions.

\subsection{Related Results}

Practical techniques for physical layer cryptography can be roughly classified into three categories according to where they are implemented in the physical layer transmission chain: at channel, signal and coding level. Techniques at channel level use the shared random channel for distilling keys (which can be used by traditional cryptography). At signal level, these techniques alter the modulation process and may require modifications in the design of transmitter and receiver. Finally, techniques at coding level are based on code design methods that differ from classic code design and are modulation and hardware independent.
 
For the terrestrial wireless channel, earliest studies \cite{Koorapaty2000} \cite{Hero2003} proposed the use of the reciprocity property of the wireless channel in the different transmission domains of time, space and frequency for secret key generation. With the advent of multiple-input multiple-output (MIMO) antennas, the same ideas were tested for wireless MIMO systems for secret cryptographic key agreement \cite{Li2005} and secrecy capacity \cite{ParadaBlahut2005,LiuShamai2009, Khisti2010, Khisti2010-1,Oggier2015}. These works usually rely on perfect CSI of the eavesdropper's channel.
The role of radio frequency (RF) channel fading for the different techniques has been well studied \cite{Barros2006}, \cite{Liang2006}, \cite{Li2006}, \cite{Gopala2008}. In \cite{Bloch2008} it is proved that, in principle, any average secure communication rate below the average secrecy capacity of the fading channel is achievable. In \cite{GoelNegi2008} a masking signal in the modulation was proposed to obfuscate the eavesdropper when demodulating the analog signal. However, the noise-like nature of the masking signal induces a large dynamic range that reduces its applicability. The use of error-control coding has also been widely investigated. Code design based on coset coding as first proposed in \cite{Wyner1975}\cite{OzaWyn1984} aims at achieving several goals simultaneously: to maximize Bob's correct decoding probability and information rate and minimize information leaked towards Eve. A large body of constructions and results exist from information theoretical perspective \cite{InfoTheoreticSec}, \cite{PhysicalLayerSecurity} and from lattice algebraic perspective \cite{Belf2011}\cite{Belf2013}\cite{Ling2014}\cite{Oggier2015}. 

For the satellite wireless channel, the potential of physical layer security has been less investigated. Physical layer security in satellite systems has been traditionally based on spread spectrum, i.e. at modulation ("waveform") level. LPI and LPD waveforms \cite{Wu2005}\cite{Wu2006}\cite{Liao2009} along with signal processing and channel coding techniques \cite{Baldi2016} are highly effective against jamming attacks and interception. Early studies of information-theoretic physical layer security focused on multibeam satellite communications. In \cite{Lei2011} joint power and antenna weights optimization at beamforming level is proposed to meet individual secrecy rate constraints. Beamforming weights are obtained from zero forcing constraints on co-channel interference among legitimate users and  eavesdroppers. Subsequent follow up works \cite{Kalantari2015}\cite{Kalantari2016} derive optimal beamforming weights under different system-level assumptions such network coding protocols. In these works, secrecy rate is the secrecy performance metric. Also at modulation level but differently to weight optimization, \cite{Tollefson2015} introduces an improved masking signal method with low computation complexity and reduced dynamic signal range. Assuming the eavesdropper has perfect knowledge of modulation scheme, encoding and frame structure, the authors transform the communication security problem into a physical security. However, this work assumes security performance in terms of bit error rate. 

%While the above results mostly refer to applications over RF carriers are of increasing relevance for different types of missions \cite{Cesarone2011}\cite{Lange2012}\cite{Samian2015}\cite{Xie2013}. 
%Optical links are inherently more secure than RF carriers and improvements in throughput are expected of around two orders of magnitude. 
%However, there can still exist thread of eavesdropping under realistic channel conditions. 
%The applicability of information theoretical security to enhance secret sharing and key agreement over optical links have been investigated in \cite{Wang2014} while outage probability due to laser-beam divergence and turbulence-induced fading is discussed in \cite{LopezM2014}. In \cite{Endo2015} on-off keying modulation secrecy capacity is investigated. 
%The authors obtain upper bounds on the decoding error probability and the leaked information. However, none of these works propose wiretap code constructions.

\subsection{Contributions}
Our contributions can be summarized as follows:
\begin{enumerate}
\item We propose a wiretap channel model for the finite-length regime. The model includes an stochastic wiretap encoding method using existing practical error correcting codes. Secrecy is provided with privacy amplification, for which the finite-length secrecy metric is given that upper bounds semantic secrecy. 
\item We derive a novel RF (broadcast) satellite wiretap channel model that parameterizes stochastic degradation in angular and radial coordinates for the realistic assumption of rotational symmetry of antenna patterns. The model allows to identify stochastic degraded spatial areas on the broadcasting space region around the legitimate channel based on worst system-level assumptions
\item We propose a methodology for the design of a secure satellite physical layer and finite-length performance evaluation using existing hash codes and error correcting codes. We define as \emph{sacrifice rate} the fixed fraction of the overall coding rate budget for reliability that needs to be allocated to ensure secrecy. We illustrate our proposed design method with numerical results using practical error correcting codes in current standards of satellite communication. 

\end{enumerate}

This work is structured as follows. Section II our proposed wiretap channel model for the finite-length regime, wiretap code construction method and secrecy metric. Section III derives the satellite wiretap channel model for RF links and the stochastically degraded spatial regions. Section IV presents the application for the design of a secure satellite physical layer and finite-length performance evaluation. A discussion for realistic scenarios is presented in Section VI and conclusions and further work are outlined in Section VII.

\section{Wiretap Channel Model for Finite-length Regime}
\label{sec:security-criteria}

\subsection{Classic Wiretap Channel Model for the Infinite Regime}

Shannon introduced the classic model of a cryptosystem in 1949, where Eve has access to an identical copy of the cyphertext that Alice sends to Bob. Shannon defined perfect secrecy to be the case when the plaintext and the cyphertext are statistically independent. Perfect secrecy is motivated by error-free transmission and requires that Bob and Alice share a secret key. 

A. Wyner in 1975~\cite{Wyner1975} relaxed the stringent condition of perfect statistical independency of Shannon's cryptosystem and introduced the wiretap channel model. In this model, transmission is not error-free.  The wiretap channel model is composed of two channels. Consider the message set ${\cal M}_n=\{1, 2, ..., M_n\}$. The channel from the legitimate transmitter (Alice) to the legitimate receiver (Bob) is referred to as the "main'' channel, and is considered to be a memoryless channel characterized by input alphabet ${\cal X}$, output alphabet ${\cal Y}$, a transition probability $W_{Y |X}$. The other channel from Alice to a passive adversary (Eve) is referred to as the "eavesdropper's channel'', and consists of another memoryless channel characterized by input alphabet ${\cal X}$,  output alphabet ${\cal Z}$, and transition probability $W_{Z|X}$. Finite-length sequences of the random variables $X$, $Y$, $Z$ are denoted as $X^n$, $Y^n$, $Z^n$ and the corresponding sets ${\cal X}_n$, ${\cal Y}_n$, ${\cal Z}_n$. This model supposes that the statistics of both channels are known to all parties, and that authentication is already done\footnote{A small secret key is needed to authenticate the communication. It is known that $\log n$ bits of secret are sufficient to authenticate $n$ bits of data~\cite{Wegman1981}}. Also, it is assumed that Eve knows the coding scheme used by Alice and Bob. This probabilistic channel model is known as Type I and there also exists a combinatorial channel model, which is known as Type II \cite{OzaWyn1984}. It is given as an adversarial channel model, where the intruder is allowed to observe $\mu\leq n$ components of the codeword. For either channel model, the classic wiretap code design goal is the simultaneous provision of reliability and security and requires stochastic encoders. In~\cite{Wyner1975} Wyner considered the special case where both the main and the eavesdropper channel are discrete memoryless channels (DMCs). Further, under the assumption that the eavesdropper channel is stochastically degraded with respect to the main channel, he defined and obtained a unique parameter characterizing the wiretap system, the secrecy capacity. This parameter means that for $\epsilon > 0$, there exist coding schemes that can provide secure and reliable rate (secrecy rate) $R_{s}>C_{s}-\epsilon$. Wyner proved that Alice can send information to Bob in perfect secrecy over a noisy channel without sharing a secret key with Bob initially. 
\begin{figure}[tbh]
\centering
\includegraphics[scale=0.35]{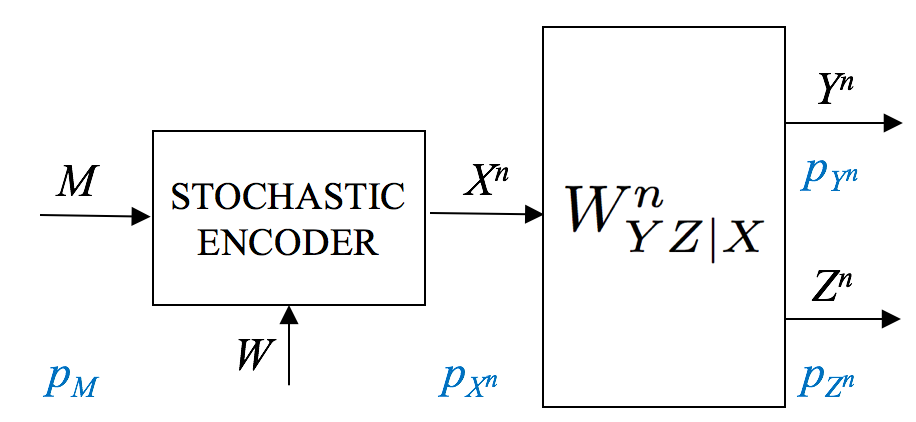}
\protect\caption{The wiretap channel model.}
\label{Fig1}
\end{figure}
I.~Csisz\'ar and J.~K\"orner generalized in 1978~\cite{Csiszar1978} Wyner's wiretap channel model by considering the information-theoretic discrete memoryless broadcast channel as shown in Fig. 1.The randomizer variable $W$ is uniformly distributed and is independent of $M$. The noisy broadcast channel is characterized by the conditional distribution $W_{YZ |X}$ so that Wyner's model corresponds to the special case where $V\rightarrow X\rightarrow YZ$ is a Markov chain and $P_{YZ |X}$  factors as $W_{Y |X}W_{Z |X}$. The secrecy capacity $C_{s}$ in this case is the maximal rate at which Alice can reliably send information to Bob such that the rate at which Eve obtains this information is arbitrarily small. It is positive whenever the legitimate channel has an advantage in terms of the broadcast channel's conditional distribution  $W_{YZ |X}$. In this case, the secrecy capacity of a discrete memoryless wiretap channel is completely characterized as~
\begin{align}
  \label{eq:secrecy_capacity}
  C_s = \max_{p_{VX}}\left(I(V;Y)-I(V;Z)\right)_+,
\end{align}
where
$I(V;Y)$ and $I(V;Z)$ 
express the mutual informations under the distributions
$p_{VXY}:=W_{Y|X}p_{XV}$
and $p_{VXZ}:=W_{Z|X}p_{XV}$,
and 
$(a)_+$ is $a$ for a positive number $a$ and is $0$ for a negative number $a$. 
The use of channel prefixing by using additional $V$ can be used in case $I(X; Y ) \leq I(X;Z)$ for all input distributions \cite{Csiszar1978}
but also as a technique to handle system-related cost constraints \cite{Han2014}, in which case the auxiliary channel $p_{X|V}$ is optimized so that the output sequence satisfies some target cost.

\subsection{Proposed Wiretap Channel Model for the Finite-length Regime} 

The classic wiretap model as first proposed by Wyner and then generalised by I.~Csisz\'ar and J.~K\"orner was later strengthended to meet cryptographic security standards in~\cite{Maurer1993} and more recently in~\cite{Bellare2012}\cite{Bellare2012arXiv}. In doing so, logical equivalences are found between existing cryptographic security measures and classic information theoretic security metrics. The cryptographic approach to the wiretap channel can also be framed within spectrum information-theoretic methods, as established by Hayashi \cite{hay-wire}. This is the approach we adopt here where a stochastic process of encoding and decoding is assumed using the privacy amplification method. Then, the chosen secrecy metric meets cryptographic criteria and is given in the finite length regime. Our proposed wiretap channel model for the finite-length regime is given as follows.
\subsubsection{Stochastic wiretap encoder} 
The stochastic wiretap encoder is based on  the privacy amplification method REFS. This method decouples reliability and secrecy, enabling the implementation of different security protocols. The code uses hash functions to approximate the statistical random process $Z^n$ as induced by a uniform input distribution nearly identical to a uniform $p_{X^n}$ therefore not requiring uniform message distribution as
  \begin{align}
  q_{Z^n} = \prod_{i=1}^{n}  \sum_{x \in {\cal X}}^{n} W_{Z|X}(z|x) q_{X}(x).
  \end{align}
This model is shown in Fig. 2. Note that it is a particular design of the wiretap model in Fig. 1. 
\begin{figure}[tbh]
\centering
\includegraphics[scale=0.4]{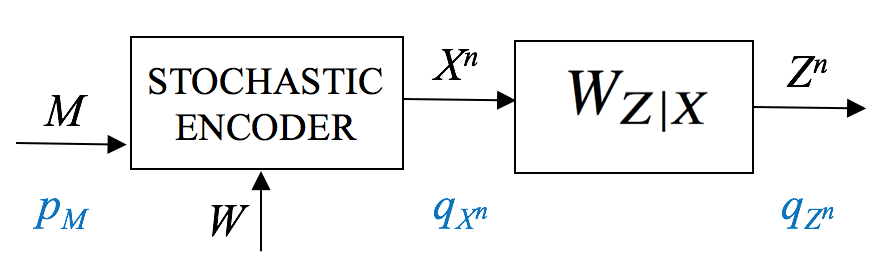}
\protect\caption{The wiretap channel model for Eve in this paper.}
\label{Fig2}
\end{figure}
 
The practical construction of a wiretap code with this method is as follows. We can make a wiretap code from a conventional  linear error correcting code. e.g., a capacity achieving LDPC or Polar code, which has efficient encoding and decoding at finite length. In this construction, assuming $n$ uses of the channel and a target of $k_n$ secure bits to be transmitted, we need to sacrifice $k_n'$ bits by attaching randomized hash function $F$ \cite{Wegman1981,Hayashi-Tsurumaru}. That is, to transmit $k_n$ bits securely, we need to employ a linear error correcting code that can transmit $k_n+k_n'$ bits. Let's denote the error correcting coding rate at finite length as $\rho(n,\epsilon^B_n)$, where $\epsilon^B_n$ is the reliability guaranteed to Bob (measured as average error probability) with finite coding size $n$. Let's also denote the coding rate to be sacrificed, the {\em sacrificed rate}, as $\rho_{{\rm sac}}(n, \delta^E_n)=k_n'/n$, where $\delta^E_n$ is the secrecy guaranteed against eavesdropping (measured with secrecy metric introduced later) with finite coding size $n$. Then, the \emph{secure coding rate}, $\rho_s$, is limited by the sacrificed rate as
\begin{align}
\rho_s(n, \epsilon^B_n, \delta^E_n) = \rho(n,\epsilon^B_n) - \rho_{{\rm sac}}(n, \delta^E_n).\label{eq:SecRate}
\end{align}
\subsubsection{Secrecy metric in the finite length regime} 
Several metrics have been introduced in the wiretap literature to relax the total independency required by Shannon perfect secrecy. The metric introduced by Wyner in \cite{Wyner1975} was later called \emph{weak secrecy} and is given as the normalized mutual information rate
under the assumption of uniform distribution $p_M$ \cite{TDCMJ}\cite{MV}. The weak security metric criterion was strengthened to $\emph{strong secrecy}$ by subsequent improvement of Wyner's model in \cite{Maurer1994} and coincides with the mutual information also with uniform distribution $p_M$\cite[Appendix D-C]{Hayashi2012}\cite{CBK} and is given as
 \begin{align}
    \label{eq:strong_secrecy}
    \text{S}_\text{strong}({M};Z^n)
={\mathbb{I}}(M;Z^n) = 
{\mathbb{D}}(p_{{M}Z^n} \Vert p_{M}p_{Z^n}).
  \end{align}
 Since message $M$ does not necessarily obey the uniform distribution, in general, the information theoretic secrecy criterion was formalized by Bellare, Tessaro, and Vardy~\cite{Bellare2012,Bellare2012arXiv} by adapting the notion of semantic secrecy used in computational cryptography~\cite{Goldwasser1984} while a similar criterion has been introduced in the security analysis of quantum key distribution \cite[(7),(9)]{QKD2007}. 

The secrecy metric used in this paper is an upper bound of the semantic secrecy metric. Instead of a quantity that gives the resulting semantic security we have an upper bound as follows \cite{Bloch2015}
   \begin{align} 
        &\text{S}_\text{strong}({M};Z^n) = {\mathbb{D}}(p_{{M}Z^n} \Vert p_{M}p_{Z^n})\nonumber \\
        =& {\mathbb{D}}(p_{{M}Z^n} \Vert p_{M}q_{Z^n}) - {\mathbb{D}}(p_{Z^n} \Vert q_{Z^n})\nonumber \\
        \leq & {\mathbb{D}}(p_{{M}Z^n} \Vert p_{M}q_{Z^n})\nonumber \\
        =& {\mathbb{E}}({\mathbb{D}}(p_{Z^n|M} \Vert q_{Z^n})|M)\nonumber
  \end{align}

Since inducing a uniform distribution obeys to $\max_{p_{M}}\text{S}_{\text{strong}}({M};Z^n)$, semantic secrecy can be guaranteed by upper bounding the quantity $\max_{p_{M}}\text{S}_{\text{strong}}({M};Z^n)$ with a proper coefficient. Since the security criterion $\max_{p_{M}}\text{S}_{\text{strong}}({M};Z^n)$ expresses the strong security independent of the source distribution $P_{M}$, it is called the \emph{source universal security} \cite[Section XIII]{Hayashi2012}. Another related measure is the \emph{effective secrecy} in \cite{Han2014}, which refers to measures in \cite[Section III]{hay-wire}. 

When we employ the random coding for the reliability
and the resolvability for the secrecy,
both metrics $\epsilon_n^B$ and $\delta_n^E$ in the finite-length (i.e. non-asymptotic) regime are exponential functions. The analytical expression for the former has been derived in \cite{Gallager} and for the latter has been derived in \cite{hay-wire}\cite{Hayashi2012}. 

In this paper, we address a practical construction of wire-tap code by using existing error correcting code
and a randomized hash function $F$ \cite{Hayashi2011}.
While the detail construction is given in Appendix A,
the calculation complexity of its implementation is not so large.
When we employ this wire-tap code,
assuming $W_{Z|X}$ is symmetric, as explained in Appendix B,
we can evaluate our considered metric in the finite-length as the average with respect to the hash function $F$. 
That is, it is upper bounded as
\begin{align}
{\mathbb{E}}_F  \text{S}_{\text{strong}}(s|M;Z^n)
\le \frac{1}{s} 2^{-s k_n'}e^{n E_{0}(s|W_{Z|X},p_{X,U})}. \label{H12-13B}
\end{align}
for $s \in [0,1]$, where

\begin{align*}
E_{0}(s|W_{Z|X},p_{X,U}) &= E_{0,\max}(s|W_{Z|X})\\
&= \log \int (\sum_{x \in \{1,-1\}} 
\frac{1}{2} W_{Z|X}(z|x)^{\frac{1}{1-s}} )^{1-s} dz .
\end{align*}

\section{RF (broadcast) wiretap satellite channel model}

\subsection{Channel model}
We consider a typical satellite RF channel with $2^m$-level phase shift keying ($2^m$-PSK) modulation with additive white Gaussian (AWGN) noise. We assume the realistic case of space links with no fading. The complex low pass equivalent signal is
\[
X_{i}=\exp\left[{j\frac{2\pi\left(i-1\right)}{2^m}}\right],\begin{array}{cc}
 & i=1,\ldots,2^m\end{array}.
\]
We assume the symbol duration, $T_{s}$, and \emph{unitary energy} constellation $E_s = E\left[\left|X_{i}\right|^{2}\right]=1$. 

For $m = 1$ we have the binary signal called BPSK. We can represent the BPSK signal as two points located on a single geometrical basis (orthonormal carrier) with one point located at $+\sqrt{E_b}$ and another point located at  $-\sqrt{E_b}$ with $E_bR_b=E_b/T_b$, $T_b$ is the bit period and $R_b$ is the bit rate. 

%\begin{figure}[tbh]
%\centering
%\includegraphics[scale=0.3]{RhosAngles1}
%\protect\caption{Visualization in 3D of the scenario showing Alice, Bob and Eve w.r.t. antenna radiation patterns (assumed for simplicity with rotational symmetry). }
%\label{fig:Scenario3D}
%\end{figure}
\begin{figure}[tbh]
\centering
\includegraphics[scale=0.3]{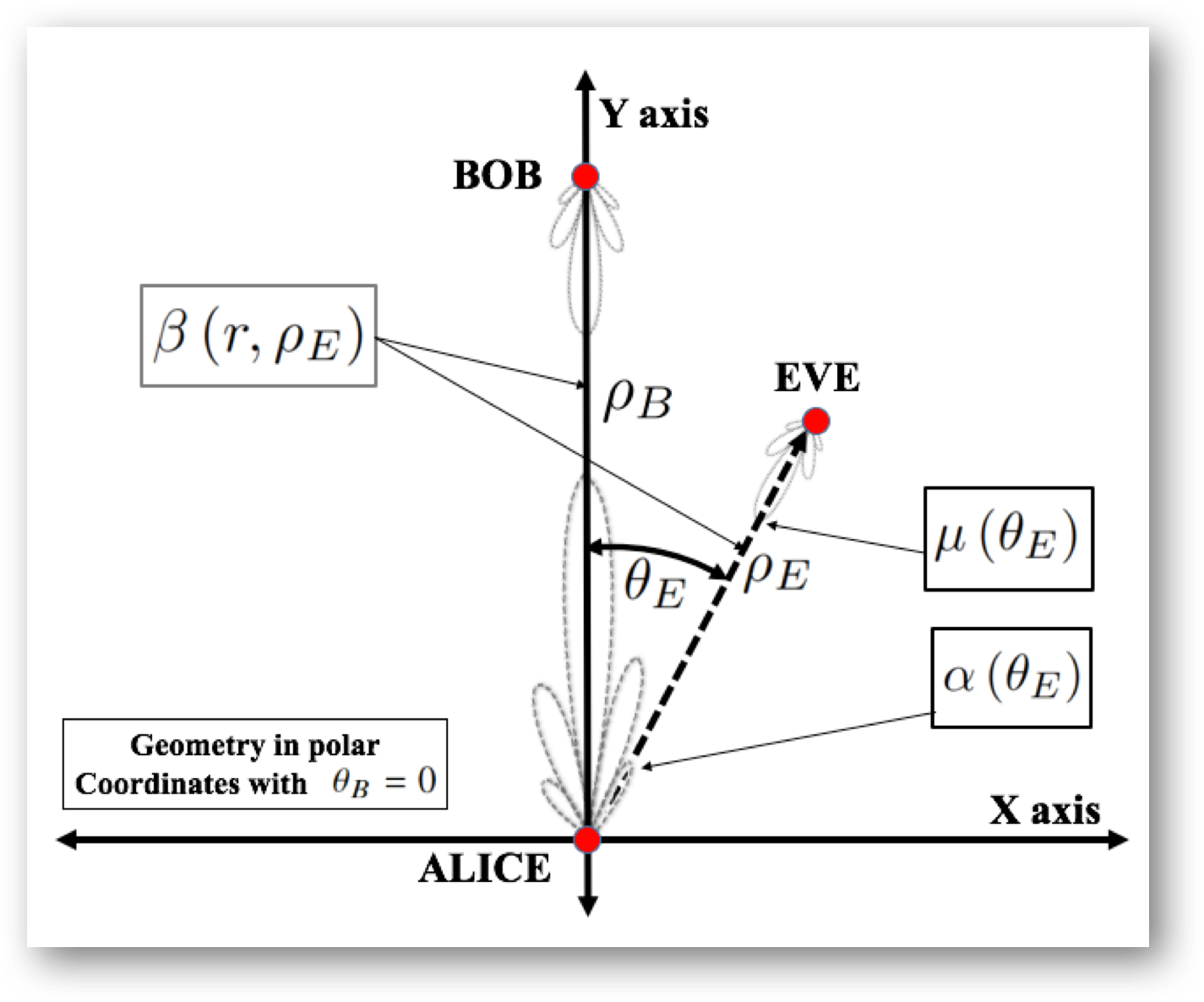}
\protect\caption{Under the (realistic) assumption of rotational symmetry of the antenna patterns, polar coordinates $\rho_B$, $\rho_E$, and $\theta_E$ (with $\theta_B=0)$ in 2D capture the 3D scenario of spatial relative positions of Alice, Bob and Eve w.r.t. antenna boresight.}
\label{fig:Scenario2D}
\end{figure}

Denote as  $h_{Y}\left(\pi_{B},\lambda_{c}\right)$ and $h_{Z}\left(\pi_{E},\lambda_{c}\right)$
the deterministic coefficients that describe the (square root of)
the power decay due to propagation. Both coefficients can be functions of time for moving Eve and Bob. 
%Fig. \ref{fig:Scenario3D} shows a 3D snapshot at a point in time of our scenario. 
Under the (realistic) assumption of rotational symmetry of the antenna patterns, we can use polar coordinates to represent the scenario such as $\pi_{B}=\left(\rho_{B},\theta_{B}\right)$ and $\pi_{E}=\left(\rho_{E},\theta_{E}\right)$ describe Bob's and
Eve's polar coordinates. Alice's location is set as center of the wiretap satellite channel coordinate system. Bob's radial coordinate is $\rho_{B}$ and Bob's angular coordinate is set as reference of the polar angles, i.e.  $\theta_{B}=0$ as shown in Fig. \ref{fig:Scenario2D}. $\lambda_{c}$ is the transmission wavelength, i.e. $\lambda_{c}=c/f_{c}$ with $c$ the speed of light. 

For simplicity, we now assume unitary transmission power. Therefore, the large-scale signal models of the received signals after demodulation during a symbol interval are
\begin{equation}\label{H21}
\begin{split}
Y%\left(\pi_{B},\lambda_{c}\right)
&=h_{Y}\left(\pi_{B},\lambda_{c}\right)X+N_{B};\\
Z%\left(\pi_{E},\lambda_{c}\right)
&=h_{Z}\left(\pi_{E},\lambda_{c}\right)X+N_{E};
\end{split}
\end{equation}

The dependency of the propagation attenuation with the wavelength is only due to how the system parameters are measured for computing the free loss propagation. The terms $N_{B}$ and $N_{E}$ are complex circular Gaussian random variables with zero mean and noise energy per (complex) symbol at Bob's receiver $n_{B}'$ and $n_{E}'$, respectively.

For the purpose of security analysis methodology, without loss of generality it is sufficient to consider a simplified e.g. uplink budget between Alice and Bob as in Fig. \ref{fig:Scenario2D}. In this case, the deterministic coefficient for Bob can be expressed as 
\begin{align}
h_{Y}\left(\pi_{B},\lambda_c\right)=\sqrt{g_{A,max}g_{B,max}}\frac{1}{\rho_{B}}\left(\frac{\lambda_{c}}{4\pi}\right),
\end{align}
where $g_{A,max}$ and $g_{B,max}$ are the gains of Alice's antenna
and Bob's antennas to each other direction, so antennas are aligned. We assume perfect polarization
antenna matching and we omit additional transmission/reception path
losses and (geo-climatic) atmospheric contributions as a first approximation.
For reliable communication, the satellite system design should ensure
a minimal received power level at the legitimate receiver so that
the link budget is closed. 

Let's now introduce the parameter $\alpha\left(\theta_{E}\right)$
to account for spatial attenuation due to Alice's antenna radiation's
pattern with respect to the Bob's boresight angle. $\alpha\left(\theta_{E}\right)$ can be considered exactly in case the antenna pattern is known, or 
otherwise it can be considered in terms of the allowed emission of radiation according to space regulations. Denote Eve's antenna gain as $g(\theta_E)$ and introduce the parameter $\mu\left(\theta_{E}\right)$ to
account for the relative antenna gain between Bob and Eve, i.e. $\sqrt{g_{E}\left(\theta_{E}\right)}=\mu\left(\theta_{E}\right)\sqrt{g_{B,max}}$ with 
\begin{align}\label{H32}
\mu_{min} \leq \mu\left(\theta_{E}\right) \leq \mu_{max}.
\end{align}
Note that we consider that Eve's antenna can be either better or worse than Bob's. The angular dependency may be given according to general assumptions on Eve but it can also be dropped by assuming a worst case value for the scenario under analysis, i.e. $ \mu_{min} \leq \mu \leq \mu_{max}$. 
We also define $\beta\left(r,\rho_{E}\right)$ to account for relative propagation losses between Bob and Eve as
\begin{align}\label{H33}
\beta^2\left(r,\rho_{E}\right)=\frac{\rho_{B}^{2}}{\rho_{E}^{r}}.
\end{align}
The exponent $r$ accounts for the power attenuation decay that affects Eve's propagation channel. Different values of the exponent model
correspond to different assumptions about Eve. Specifically, Eve can be modeled as a terrestrial, aerial or satellite station. For example, while for the satellite case $r=2$, in case of aerial Eve, a good assumption is to consider a large scale two-ray ground multipath model,with $r>2$. It is also known that for unmanned vehicles at low elevation, the propagation law has strong dependency with height and this can also be captured by $r$. 
Finally let's define a coefficient $\gamma_n$ such that 
\begin{align}
\gamma_n^{min} \leq \gamma_n \leq \gamma_n^{max}
\label{H34}
\end{align}
to account for the relative performance of Eve's w.r.t. Bob's receivers in terms of additive white Gaussian noise (AWGN) power. Note that we consider that Eve's receiver can be either better or worse than Bob's. With these parameters we can then write 
\begin{align}
h_{Z}\left(\pi_{E},\lambda_c\right)=h_{Y}\left(\pi_{B},\lambda\right)\alpha\left(\theta_{E}\right)\mu\left(\theta_{E}\right)\beta\left(r,\rho_{E}\right).\label{hlz}
\end{align}

Rewriting $h_{Y}\left(\pi_{B},\lambda_{c}\right)X$ to $X$,
we can simplify \eqref{H21} as
%???Dividing by the common path loss channel coefficient in Bob's and Eve's deterministic path losses, we can re-write the model so that Bob's signal is the reference for Eve's signal
\begin{equation}\label{eq:Gaussig2}
\begin{split}
Y&=X+\sqrt{n_{B}}X_1;\\
Z&=\gamma_gX+\sqrt{\gamma_nn_{B}}X_2;
\end{split}
\end{equation}
with $X_1$ and $X_2$, zero-mean circular complex Gaussian random variables with unit variance and the deterministic coefficient $\gamma_g $ is defined as
\begin{align}
\gamma_g(\theta_E, \rho_E, r): = \alpha\left(\theta_{E}\right)
\mu\left(\theta_{E}\right)\beta\left(r,\rho_{E}\right).\label{grrt}
\end{align}
%The concrete forms of $\mu\left(\theta_{E}\right)$ and $\beta\left(r,\rho_{E}\right)$
%are given in Supplement while $\alpha\left(\theta_{E}\right)$ will be given later.
The model shows the two components of physical degradation: geometrical captured by $\beta\left(r,\rho_{E}\right)$ and system-related captured by $\alpha\left(\theta_{E}\right)$, which is controlled by Alice and $\mu\left(\theta_{E}\right)$, which is controlled by Eve. Overall, the channel parameters are summarized in Table \ref{T1}.

\begin{table}[htpb]
  \caption{Summary of parameters in RF satellite wiretap channel model}
\label{T1}
\begin{center}
  \begin{tabular}{|c|l|c|}
\hline
Parameter & Explanation & Eq. etc.\\
\hline
$\theta_E$ &Angle between Bob and Eve. &Fig. \ref{fig:Scenario2D}\\
\hline
$\rho_B$ &Distance to Bob. &Fig. \ref{fig:Scenario2D}\\
\hline
$\rho_E$ &Distance to Eve. &Fig. \ref{fig:Scenario2D}\\
\hline
{$n_B$} & Noise power level at Bob's receiver.& \eqref{eq:Gaussig2}\\
\hline
\multirow{3}{*}{$\gamma_g(\theta_E, \rho_E, r)$} & Deterministic channel attenuation &
\multirow{3}{*}{\eqref{grrt}}\\
&coefficient.  It captures Eve's channel  & \\
&degradation w.r.t. Bob's channel.  &\\
\hline
\multirow{2}{*}{$\gamma_n $} & 
Ratio between the noise power levels at 
  & \eqref{H34}\\
&EVe's and Bob's receivers.
& \eqref{eq:Gaussig2} \\
\hline
\multirow{3}{*}{ $\alpha(\theta_B)$}  & 
Spatial attenuation due to Alice's antenna & \multirow{3}{*}{\eqref{H31}}\\
&radiation pattern. It can be given as  &\\
&antenna gain or allowed radiation mask.& \\
\hline
\multirow{5}{*}{ $\mu(\theta_E)$} & Eve's antenna gain. For convenience is &
\multirow{5}{*}{\eqref{H32}}\\
& taking as $\sqrt{g_{E}\left(\theta_{E}\right)}=\mu\left(\theta_{E}\right)\sqrt{g_{B,max}}$ 
& \\ 
& with $\mu(\theta_{E})$ taking values between 0 and &\\ 
& 1.  For convenience, in this paper, we &\\
& assume $\mu_{min} \leq \mu \leq \mu_{max}$.&\\ 
\hline
\multirow{4}{*}{ $\beta(r,\rho_E)$} & Coefficient relating Bob's and Eve's path 
& \multirow{4}{*}{\eqref{H33}}\\
&losses. It considers Friis model for Bob   &\\
&and Eve. For Eve, a power attenuation   & \\
&decay of ``$r$'' instead of 2 is considered.&\\
\hline
  \end{tabular}
\end{center}
%\par\vspace{3ex}\par
\end{table}

\subsection{Stochastically degraded spatial regions}

As discussed in Section II.A, the secrecy capacity of a memoryless wiretap channel in the infinite-length regime is completely characterized as \eqref{eq:secrecy_capacity}, which is statistically described by the conditional distribution $W_{YZ|X}$. Further, under the assumption that the eavesdropper channel is stochastically degraded with respect to the main channel, the secrecy capacity is positive. Hence, without discussing the block-length and the choice of error correcting code for the finite-length construction in (\ref{eq:SecRate}), the condition for positive secrecy capacity is a necessary condition for positive secrecy rate. 

We now use our RF wiretap satellite channel model to identify the stochastically degraded spatial regions. For this, recall that \cite{CoverThomas} physically degraded and stochastically degraded channels belong to the same equivalence class of broadcast channels, having
the same conditional marginals and hence same capacity. Denote the signal-to-noise ratio (SNR) for Bob as $\eta^B= \frac{E_s}{n_B}$, the wiretap channel model for Eve describes all space around the satellite channel in polar coordinates, hence physical and stochastic degradation is given as
\begin{align}
\centering
\eta^B &> \frac{\gamma_g^2(\theta_E, \rho_E, r)}{\gamma_n}\eta^B
\end{align}
yielding the following condition
\begin{align}
\centering
\gamma_g(\theta_E, \rho_E, r) &< \sqrt{\gamma_n}.\label{eq:RFCondit}
\end{align}

\begin{figure}[tbh]
\centering
\includegraphics[scale=0.33]{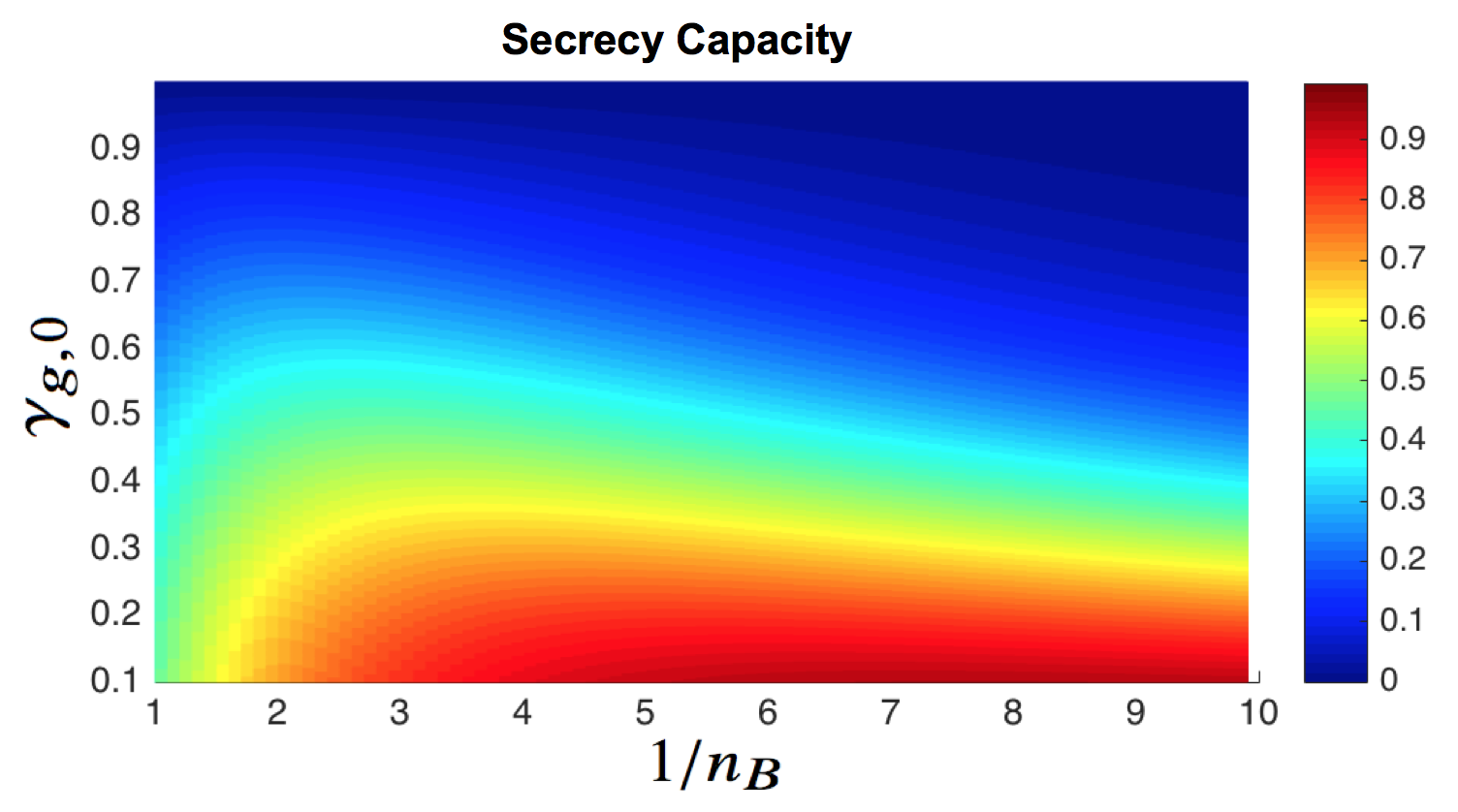}
\protect\caption{Secrecy capacity given by (\ref{H42}) represented as a function of $\gamma_{g,0}$ in the vertical axis and the signal-to-noise ratio $E_s/n_B$ for $Es=1$ in the horizontal axis.}
\label{CsBIAWGN}
\end{figure}

It is convenient to introduce the regularized parameter
\begin{equation}
\gamma_{g,0}(\theta_E, \rho_E, r) 
:= \frac{\gamma_g(\theta_E, \rho_E, r)}{\sqrt{\gamma_{n}}},
\end{equation}
%where $\gamma_{n,0}$ is the noise coefficient we consider for Eve. 
so that the condition (\ref{eq:RFCondit}) now is 
%$$\gamma_{g,0}(\theta_E, \rho_E, r) < 1,\:\:\:[\gamma_{g,0}(\theta_E, \rho_E, r)]_{dB} < 0$$
$\gamma_{g,0}(\theta_E, \rho_E, r) < 1$
and the regularized signal of Eve  is written as
\begin{align}
\frac{1}{\sqrt{\gamma_{n}}} Z= \gamma_{g,0}X+ \sqrt{n_B}X_2.\label{E12}
\end{align}
As discussed in Appendix C, once the condition (\ref{eq:RFCondit}) is satisfied,
the secrecy capacity is given as a function of $\gamma_{g,0}$ and $n_B$
in the following way in the BPSK case.
\begin{align}
&C_s(\gamma_{g,0} , n_B) \nonumber\\
=&\int_{-\infty}^\infty  \frac{1}{\sqrt{8\pi n_B}}u\Big[e^\frac{-(y+1)^2}{n_B} 
+ e^\frac{-(y-1)^2}{n_B} \Big]dy\nonumber\\
&-\int_{-\infty}^\infty  \frac{1}{\sqrt{8\pi n_B}}u\Big[e^\frac{-(z+\gamma_{g,0})^2}{n_B} 
+ e^\frac{-(z-\gamma_{g,0})^2}{n_B} \Big]dz.
\label{H42}
\end{align}
where $u(x):= - x \log x$. Then, we have a numerical calculation of the secrecy capacity as a function of
$\gamma_{g,0}$ and $n_B$ as Fig. \ref{CsBIAWGN}.
Hence, it is important to characterize 
the regularized parameter $\gamma_{g,0}(\theta_E, \rho_E, r) $
as a function of $\theta_E$ as well as $\rho_E$ and $r$.

The \emph{stochastically degraded spatial regions} for RF wiretap satellite channel are then identified as
\begin{align}
\mathcal{R}_{\eta^B}^{RF}(\rho_E, r)&= \left\{ \theta_E | \gamma_{g,0}(\theta_E, \rho_E, r) <  1\right\}\nonumber \\
&= \left\{ \theta_E | \gamma_{g}^{dB}(\theta_E, \rho_E, r) - \gamma_n^{dB}< 0 \right\},\nonumber \\ \label{eq:DegradRegions}
\end{align}
%\begin{align}
%\mathcal{R}_{\eta^B}^{RF}(\rho_E, r)&= \left\{ \theta_E | \gamma_{g}(\theta_E, \rho_E, r) <  \sqrt{\gamma_n}\right\}\nonumber \\
%&= \left\{ \theta_E | [\gamma_{g}(\theta_E, \rho_E, r)] < [\sqrt{\gamma_n}] \right\}\nonumber \\
%&=\left\{ \theta_E | [\alpha(\theta_E)] + [\mu] 
%+ [\beta(\rho_E, r)] < [\sqrt{\gamma_n}] \right\},
%\end{align}
where superscript $dB$ indicates values in decibels. Hence with no assumption on Eve's location we have
\begin{align}
\mathcal{R}_{\eta^B}^{RF}= \left\{ \mathcal{R}^{RF}(\rho_E, r), \forall (\rho_E, r) \right\}
\end{align}
since it is possible to associate a propagation law depending on Eve's location. 

\begin{figure}[tbh]
%\centering
\includegraphics[scale=0.48]{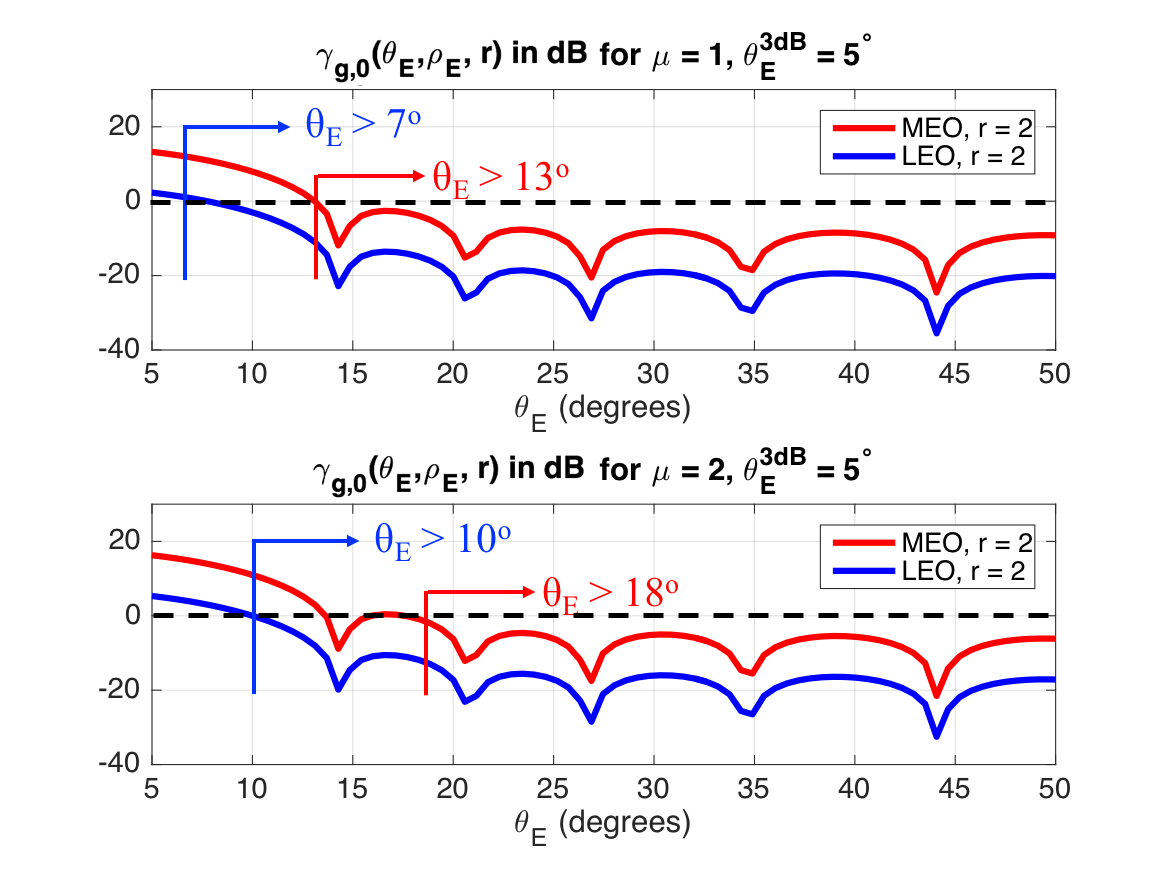}
\protect\caption{Angles defining stochastically degraded regions, $\mathcal{R}^{RF}(\rho_E, r)$, when Bob is a GEO satellite and Eve uses LEO and MEO satellites for $\theta^{3dB}_E = 5^{\circ}$ with $\gamma_n^{dB} = 0$.}
\label{fig:GEOresults1}
\end{figure}

\begin{figure}[tbh]
%\centering
\includegraphics[scale=0.48]{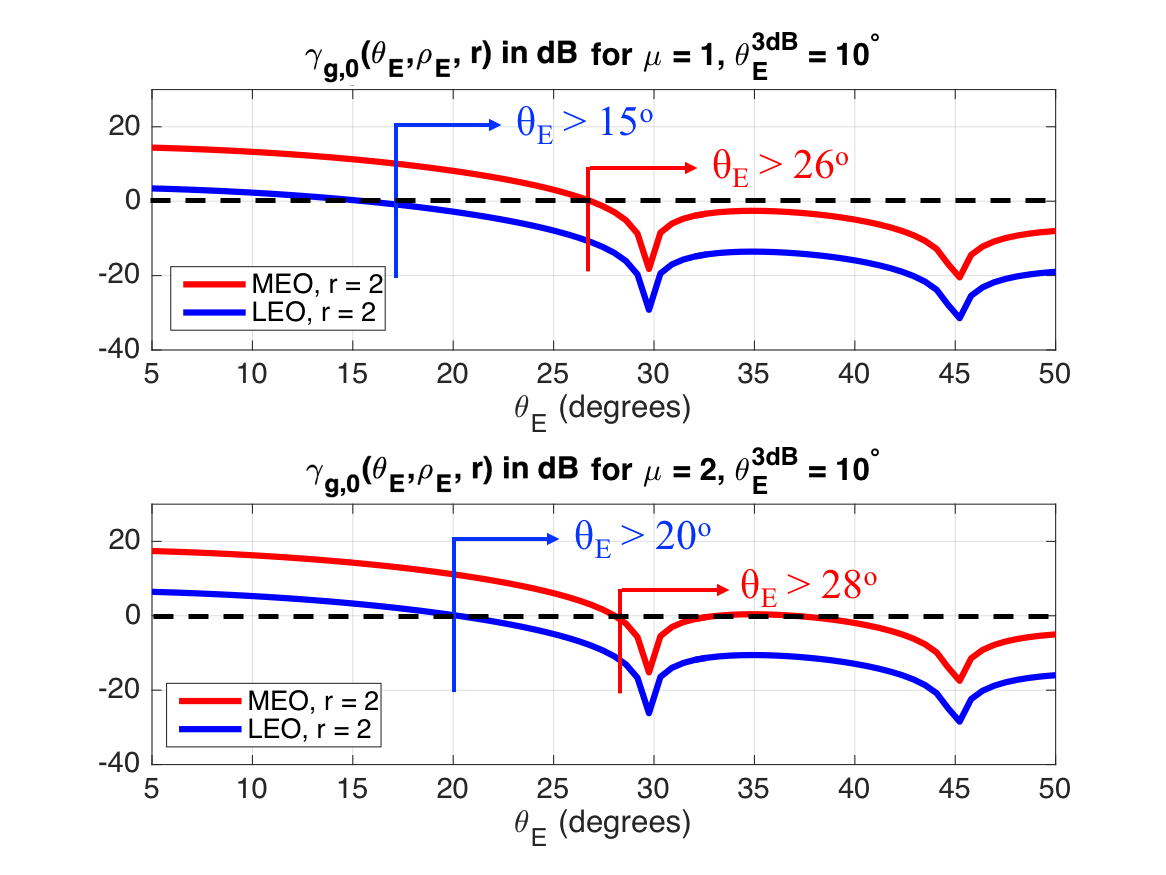}
\protect\caption{Angles defining stochastically degraded regions, $\mathcal{R}^{RF}(\rho_E, r)$, when Bob is a GEO satellite and Eve uses LEO and MEO satellites for $\theta^{3dB}_E = 10^{\circ}$ with $\gamma_n^{dB} = 0$.}
\label{fig:GEOresults2}
\end{figure}

\begin{figure}[tbh]
\centering
\includegraphics[scale=0.45]{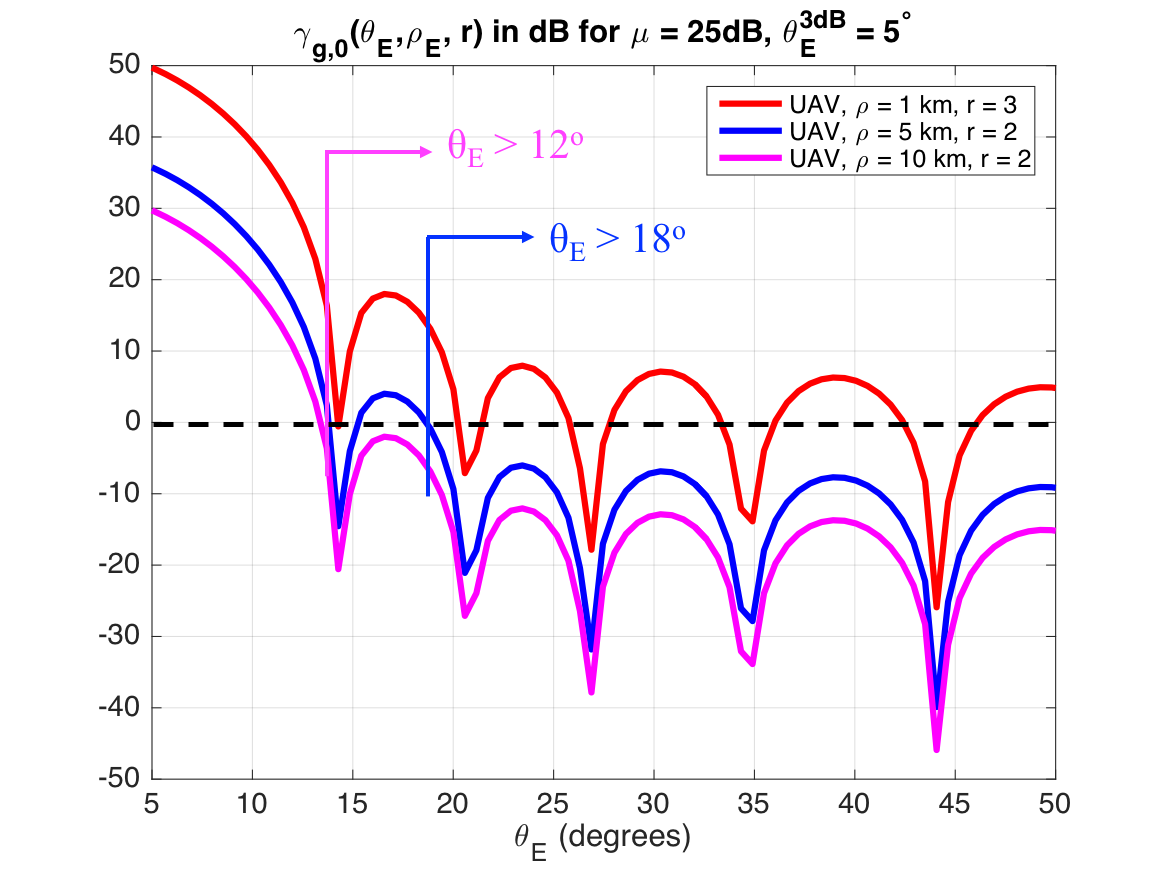}
\protect\caption{Angular regions where security can be guaranteed for Bob on a MEO orbit and Eve on different types of UAVs. It shows three different heights assuming $\mu = -25$ dB and  $\gamma_{n} = 3\:(4.77\:dB)$}
\label{fig:MEOUAV}
\end{figure}

Now, we provide some numerical examples. For the sake of clarity, in order to reduce the number of variables we fix the operation point as $E_s = n_B = 1$ and use the notation $\mathcal{R}^{RF}(\rho_E, r)$ and $\mathcal{R}^{RF}$. First, let's assume the simple case of Bob as a geostationary Earth orbit (GEO) satellite and study the effect of low Earth orbit (LEO) and medium Earth orbit (MEO). In this case it is realistic to assume $\gamma_{n}=1$. As for a (normalized) antenna pattern we can use some illustrative pattern radiation diagram with side lobes such as \cite{Caini92}
\begin{equation}\label{H31}
\alpha(\theta_E) = \frac{J_1(k\sin(\theta_E))}{2k\sin(\theta_E)} + 36\frac{J_3(k\sin(\theta_E))}{(k\sin(\theta_E))^3}
\end{equation}
where $k = 2.0712/\sin(\theta^{3dB}_E)$, with $\theta^{3dB}_E$ being the one-sided half-power angular beamwidth and $J_1$ and $J_3$ are the Bessel functions of the first kind, of order one and three respectively. Note that in case the antenna pattern is not known, the analysis can be made with $\alpha(\theta_E)$ considered in terms of the allowed emission of radiation according to space regulations (see e.g. Recommendation ITU-R S.465-6). 
We assume radial distances of 15000 km and 1200 km as illustrative of the visibility windows for MEO and LEO satellite orbits, respectively. Fig. \ref{fig:GEOresults1} and \ref{fig:GEOresults2} shows the angle from which the channel is degraded. We observe that when Eve uses a LEO satellite, Eve's channel is degraded w.r.t. Bob's channel if Eve and Bob have an angular separation greater than $7^{\circ}(15^{\circ})$ in case of equally good antennas and greater than $10^{\circ}(20^{\circ})$ if Eve's antenna is 6\:\:dB better than Bob's antenna for 3\:dB angle of $5^{\circ}(10^{\circ}).$ We observe that degradation region is smaller for MEO than for LEO. This is due to the dependency of the degradation condition with the ratio $\beta(r, \rho_E)$, namely, for fixed $\rho_B$ the smaller $\beta(r, \rho_E)$ the more degraded is the channel, and clearly the ratio $\beta(r, \rho_E)$ is smaller for MEO orbit than for LEO orbit. We also observe as expected that if the antenna directivity increases, the degraded region is also increased, and the increase is larger when Eve's antenna is not better.
The above cases are illustrative, but due to the relative movement, there is limited time window for eavesdropping. Hence, we assume now the case of Bob at e.g. MEO orbit and study the degraded spatial region in the challenging scenario where Eve uses Unmanned Aerial Vehicles (UAVs). We note that propagation channel modelling for this scenario is still a research area. Moreover, there is a wide variety of UAVs that can be used by Eve depending on the scenario under analysis. For illustration we simply assume different heights with higher path loss exponent for lower heights (see e.g. \cite{Briso2017}). Fig. \ref{fig:MEOUAV} shows three different heights assuming $\mu = -25$ dB and  $\gamma_{n} = 3 \;(4.77\:dB)$. We observe in this example scenario that when Eve uses a UAV, Eve's channel is degraded w.r.t. Bob's channel if Eve and Bob have an angular separation greater than $12^{\circ}$ and $18^{\circ}$ for radial distances of  5 km and 10 km, respectively. However, for low altitude and propagation law $r=3$, Eve's channel is never degraded. Such low altitude UAVs however may not be the best choice for Eve as it can be visually detected. All numerical values are presented in Table \ref{T3}.

\begin{table}[htpb]
  \caption{Summary of parameters in numerical analysis.}
\label{T3}
\begin{center}
  \begin{tabular}{|c|c|c|c|c|c|c|} 
\hline
& $\rho_E$ & \multirow{2}{*}{$\theta_E^{3dB}$} & \multirow{2}{*}{$r$} & $\mu$ & \multirow{2}{*}{$\gamma_{n}$} & Degradation \\
& (km) && &(dB) & &  condition \\
\hline
Fig. 7 (t) & &5& & 1(0) & 0 &\\
MEO & 15000 &&2 &  & & $\theta_E>13^{\circ} $ \\
LEO & 1200 &&2 &  & & $\theta_E>7^{\circ} $ \\
\hline
Fig. 7 (b) &&5 & & 2(6) & 0 &\\
MEO & 15000 &&2 &  & & $\theta_E>18^{\circ} $ \\
LEO & 1200 &&2 &  & & $\theta_E>10^{\circ} $ \\
\hline
Fig. 8 (t) &&10 & & 2(6) & 0 &\\
MEO & 15000 &&2 &  & & $\theta_E>26^{\circ} $ \\
LEO & 1200 &&2 &  & & $\theta_E>15^{\circ} $ \\
\hline
Fig. 8 (b) &&10 & & 2(6) & 0 &\\
MEO & 15000 &&2 &  & & $\theta_E>28^{\circ} $ \\
LEO & 1200 &&2 &  & & $\theta_E>20^{\circ} $ \\
\hline
Fig. 9 &&5 & & 0.05(-25) & 3 &\\
UAV low & 1 &&3 &  & & - \\
UAV medium & 5 &&2 &  & & $\theta_E>18^{\circ} $ \\
UAV high & 10 &&2 &  & & $\theta_E>12^{\circ} $ \\
\hline
  \end{tabular}
\\
\vspace{2ex}
(t) expresses the top and (b) expresses the bottom.
\end{center}
%\par\vspace{3ex}\par
\end{table}
\section{Application to the design of a secure physical layer satellite link}
In this section we show the practical application of the secure wiretap code construction method (\ref{eq:SecRate}) and degradation regions (\ref{eq:DegradRegions}) for the design of secure physical layer for realistic satellite channels. In order to show the application, we employ existing practical error correcting codes with hashing. A hash construction is explained in \cite{AngMas17}. We assume the discrete binary input Gaussian output (BI-AWGN) channel in the finite-length regime. Hence, we choose LDPC codes using BPSK in current satellite communication standards. Namely, we consider the LDPC code for low SNR and BPSK from DVB-S2 \cite{DVB-S2} with $\rho_{{\rm LDPC}} = 1/3$ and medium and short frame coded block size, which are $n_{{\rm med}}=32400$ and $n_{{\rm short}}=16200$ bits, respectively. Note that we restrict our error correction and the hash functions to linear operations.

\begin{figure}[tbh]
\centering
\includegraphics[scale=0.34]{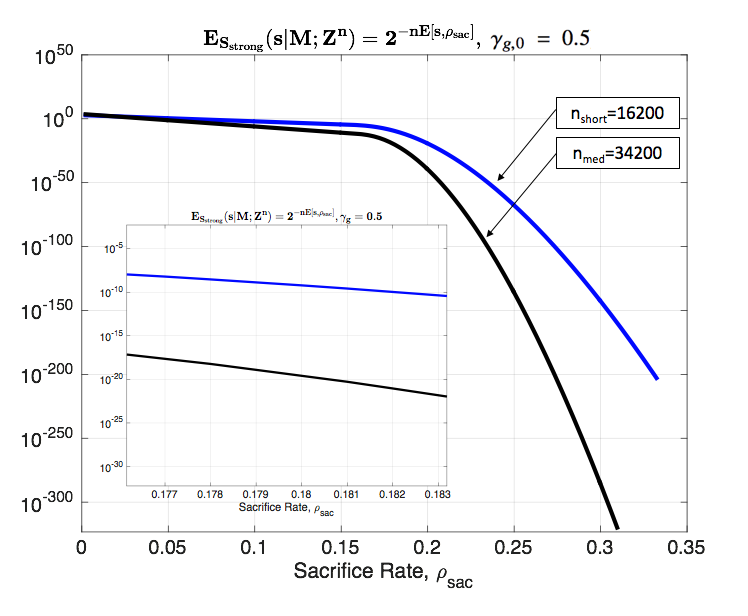}
\protect\caption{Secrecy metric as a function of the sacrifice rate (zoom in at $\rho_{{\rm sac}} = 0.18$). The evaluation is for $\gamma_{g,0}=0.5$. }
\label{fig:TradeOff}
\end{figure}

As a realistic design scenario, we tackle the case where the sacrifice rate is the design constraint i.e., for a given error correcting code length $n$, the designer \emph{fixes a fraction of the total coding rate budget for reliability} to be traded off with secrecy. 
Now, note that from a coding point of view it is of interest to evaluate the trade-off between the amount of information leakage given by the secrecy metric ($\ref{expo1}$) and the sacrifice rate design constraint, i.e. $\rho_{{\rm sac}} = x \rho_{{\rm LDPC}}$, with $x$ the fixed allocated budget to security so that $\rho_s(n, \epsilon^B_n, \delta^E_n) = (1 - x) \rho_{{\rm LDPC}}(n, \epsilon^B_n)$.\label{eq:xSecRate}
This tradeoff is illustrated in Fig. \ref{fig:TradeOff} where the security metric is shown for the two LDPC codes. We consider for clarity the secrecy metric (\ref{H12-13B}) expressed exponentially as 
\begin{align}
%{\mathbb{E}}_F  \text{S}_{\text{strong}}(s|M;Z^n) \le 2^{-n E[\rho_{{\rm sac}}(n, \delta^E_n)]} \label{expo},
{\mathbb{E}}_F  \text{S}_{\text{strong}}(s|M;Z^n)\le 2^{-n E[\rho_{{\rm sac}}]}\label{expo1},
\end{align}
with 
\begin{align}
E[\rho_{{\rm sac}}]  &= \min_{s\in[0,1]} \Bigg[s\rho_{{\rm sac}} + \frac{\log(s)}{n} - E_{0,\max}(s|W_{Z|X})\log_2(e)\Bigg].\label{expo2}
\end{align}
For example, when $x\rho_{{\rm LDPC}}=0.18$, the information leakage is lower than $10^{-5}$ for both the medium and the short frames. Higher fraction $x$ will provide higher protection. 

\begin{figure}[tbh]
\centering
\includegraphics[scale=0.27]{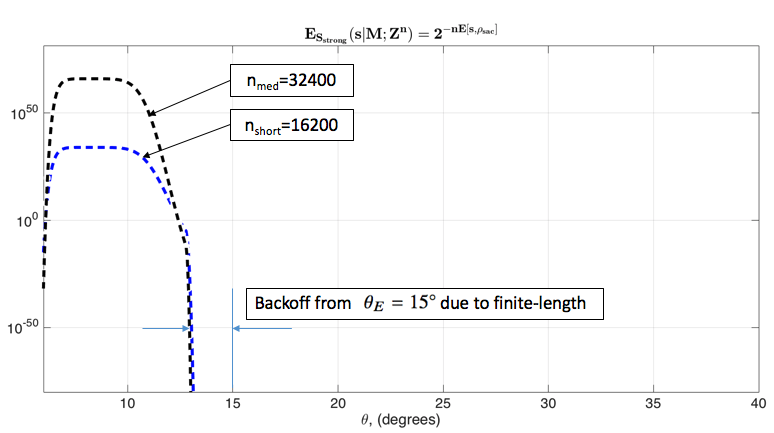}
\protect\caption{Secrecy metric as a function of the polar angle coordinate within the degraded channel region $\mathcal{R}^{RF}_{LEO} = \left\{ \theta_E |\theta_E>15^{\circ}\right\}$ (for which $E_s=n_B=1$ and $\gamma_n=1$) for design target $\rho^*_{{\rm sac}} = 0.18$.}
\label{fig:SpatialSecrecy}
\end{figure}

However, note that such tradeoff assumes a fixed (polar) angle coordinate in space for Eve for the evaluation of $E_{0,\max}(s|W_{Z|X}$) (which corresponds to some degradation for the regularized parameter 
$\gamma_{g,0}$, in Fig. \ref{fig:TradeOff} such degradation is $\gamma_{g,0}=0.5$). 
In other words, this tradeoff assumes that the designer of the secure communication system has channel side information (CSI) and system side information (SSI) of the eavesdropper. Hence, this view is not useful from a practical design point of view since the secure system design needs to provide security guarantees in all space between Eve and Bob. Hence, it is meaningful to evaluate how the information leakage occurs in the spatial regions of the channel degradation for a given choice of sacrifice rate. This leakage is what the secure communication designer needs to guarantee to the user of the secure system. 
It is however unavoidable to make assumptions about Eve's system noise receiver, parameterized by $\gamma_n$. Clearly, a reasonable design decision is to make a worst case system assumption (WSSI) on Eve's noise power. Therefore, for a given code length $n$, \emph{we fix a sacrifice rate},  $\rho^*_{{\rm sac}}=x \rho_{{\rm LDPC}}$, to be assessed and write (\ref{expo1}) as a function of the polar angle coordinate within the degraded channel region. 
For this, using the relation \eqref{grrt} for $\gamma_g$,
we do the following translation from secure information theoretical notation to secure communication system notation of our secrecy metric
$$\delta^E_n(s|\theta_{\mathcal{R}}) = {\mathbb{E}}_F  \text{S}_{\text{strong}}(s|M;Z^n), \quad \theta_{\mathcal{R}} \in \mathcal{R}_{\eta^B}^{RF}(\rho_E, r).$$
Observe that the notation now makes explicit the dependency of the security metric with physical system resources (through $\eta^B$ and $n$) and spatial degradation area (through $\rho_E$ and $\theta_{\mathcal{R}}$). Therefore we can write the exponential decay for the fixed sacrifice rate  $\rho^*_{{\rm sac}}$ as, 
\begin{align}
%{\mathbb{E}}_F  \text{S}_{\text{strong}}(s|M;Z^n) \le 2^{-n E[\rho_{{\rm sac}}(n, \delta^E_n)]} \label{expo},
\delta^E_n(s|\theta_{\mathcal{R}}) \le 2^{-n E[\theta_{\mathcal{R}}]} \label{expo3}, \quad \theta_{\mathcal{R}} \in \mathcal{R}_{\eta^B}^{RF}(\rho_E, r).
\end{align}
with 
\begin{align}
E[\theta_{\mathcal{R}}]  &= \min_{s\in[0,1]} \Bigg[s \rho^*_{{\rm sac}}+ \frac{\log(s)}{n} - E_{0,\max}(s|\theta_{\mathcal{R}})\log_2(e)\Bigg].
\end{align}
For example, for the same region $\mathcal{R}^{RF}_{LEO} = \left\{ \theta_E |\theta_E>15^{\circ}\right\}$ and design target $\rho^*_{{\rm sac}} = 0.18$, Fig. \ref{fig:SpatialSecrecy} shows the leakage decay in the angular coordinate for both $n_{{\rm med}}=32400$ and $n_{{\rm short}}=16200$ bits. The secrecy metric shows exponential decay in the stochastically degraded area. As expected, observe that stochastic degradation provides a conservative design since information leakage is already practically zero slightly below $\theta_E=15^{\circ}$.

Note that this is still a conservative approach to secure physical layer design since we evaluate the region of positive capacity (instead of the region of positive secrecy rate). However, it allows a good understanding of the problem of realistic design of a secure physical layer. This will allow to proceed further with more difficult scenarios.

\begin{figure}[tbh]
\centering
\includegraphics[scale=0.23]{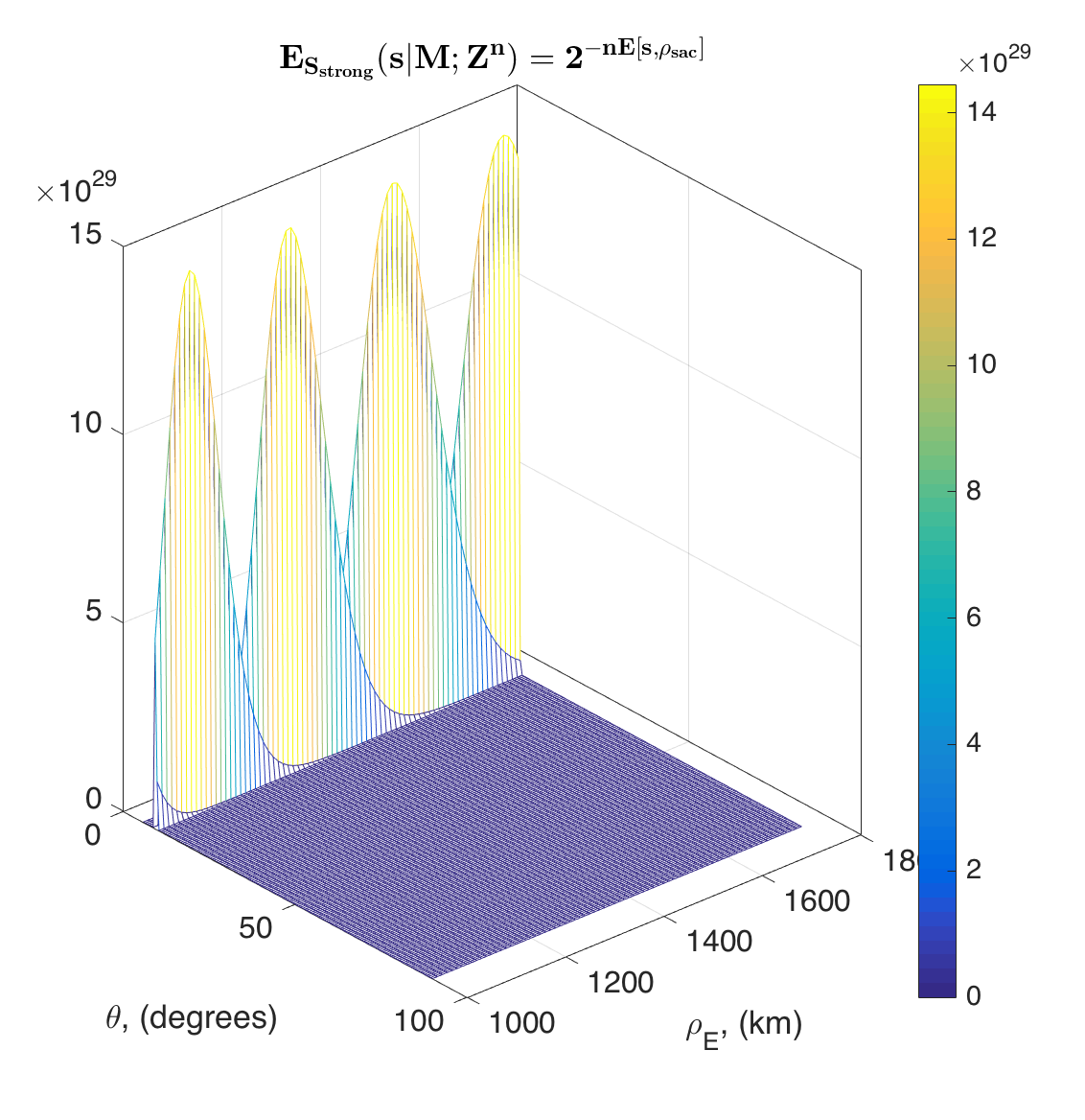}
\protect\caption{Secrecy metric in a range of polar and radial coordinates within the degraded channel region $\mathcal{R}^{RF}$ with $E_s=n_B=1$ for design target $\rho^*_{{\rm sac}} = 0.18$.}
\label{fig:SpatialSecrecy3D}
\end{figure}

Also note that while the above example is only for LEO orbit, it is straightforward to obtain the secrecy guarantees in all points of the \emph{degraded space} between Bob and Eve for any possible stochastic degradation region. As an example, Fig. \ref{fig:SpatialSecrecy3D} shows the values for a range of locations in radial coordinate. It is observed that there is no secrecy guarantee outside the degraded channel region, which is visualized as  values of the secrecy metric greater than one and the effect of the antenna still present.

Finally, note that mission-specific services may prioritize reliability over secrecy or vice versa. Also note that for certain mission-specific air interfaces, the finite-length regime needs to be analyzed considering both the link and the physical layer framing. This is due to the fact that some air interface may define very short link layer frames and/or significant physical layer overhead. In such cases, the air interface overhead could be comparable to the coding overhead.
%\subsection{Secrecy guarantees with level of secrecy as design constraint}
%In the previous section we showed a realistic design scenario in which, for a given code length $n$, the designer \emph{fixes a fraction of the total coding rate budget for reliability} to be traded off with secrecy. Now we illustrate another realistic design scenario in which, for a given code length $n$, \emph{the designer fixes a sacrifice level},  $\delta^*_n$. In this case we also need to choose some meaningful value of target secrecy. We propose to choose the value that is typically required for a satellite to be considered as quasi-error-free (QEF) channel, which is $\epsilon_n^B = 10^{-12}$, which we call quasi-eavesdropping-free (QEAF) channel, $\delta_n^* = 10^{-12}$. It is clear that a secure coding rate with higher level of guaranteed secrecy will require a higher amount of sacrificed rate for secrecy. 
%The sacrifice rate can be obtained from (\ref{expo3}) as
%\begin{align}
%\rho_{{\rm sac}}(n,\theta_{\mathcal{R}})  \le \min_{s\in[0,1]} \Bigg[\frac{E_{0,\max}(s|\theta_{\mathcal{R}})\log_2(e)}{s} - \frac{\log_2(s \delta^*_n)}{sn}\Bigg].
%\end{align}
%Therefore, $\rho_s(n, \theta_{\mathcal{R}}) = k_{LDPC}/n - \rho_{{\rm sac}}(n,  \theta_{\mathcal{R}})$.

\section{Realistic Scenario}
Finally, we discuss how we can guarantee the security in a realistic situation.
We have seen that the security parameters for transmitted message 
are upper bounded in \eqref{H12-13B}.
The upper bound given in \eqref{H12-13B} depends only on 
the parameter $\gamma_{g,0}$ due to the relation \eqref{E12}.
To guarantee the security level, we need to check whether
the parameter $\gamma_{g,0}$ belongs to a certain region.
Dependently of the values of $\gamma_{g,0}$ and the sacrifice rate $\rho_{\rm {\rm sac}}$, Fig. \ref{fig:TradeOff} guarantees the security level when 
$\gamma_{g,0}$ takes the value $0.5$.
%in $[0.1,0.9]$????. 
The numerical analysis in Section III.B clarifies when 
the degradation condition $\gamma_{g,0}< 1$ holds
dependently of $\theta_E $ with proper other parameters.
Therefore, we need to check whether the device by Eve does not exist
within the angle $\theta_E$ given in Table III.
If this condition can be adopted, we can guarantee the security information theoretically.
If we cannot deny a possibility that Eve has a stealth device, 
we need to use more powerful satellite communication, e.g., 
quantum key distribution \cite{BB84} and two-way scenario \cite{H17}.

However, when Eve has to eavesdrop so many satellite communications,
it is quite difficult for Eve to prepare such a receiver for all of satellite communications. 
In this case, Eve needs to prepare many stealth devices, which require much higher cost for Eve.
This is a serious difference from the computational security
because one powerful computer can recode the encrypted message.
Further, only a limited numbers of users can prepare such stealth device.
When we consider the cost-effectiveness, 
our assumption is reasonable for conventional user
due to the finiteness of Eve's budget.

\section{Conclusions}
In this paper we have proposed the necessary models and metrics to develop a methodology for the design of a secure satellite physical layer and finite-length performance evaluation. Our proposed method makes use of existing hash codes and error correcting codes and is based on defining the fixed fraction of the overall coding rate budget for reliability that needs to be allocated to ensure secrecy, \emph{sacrifice rate}. We illustrate our proposed design method with numerical results using practical error correcting codes in current standards of satellite communication. Our methodology does not make use of channel side information of the eavesdropper, only assumes worst case system assumptions. It allows to identify guaranteed secrecy and also where secure communication cannot be achieved, which in the satellite case corresponds to the case where the eavesdropper is situated physically near the antenna transmitter, in which case, we have provided a discussion for other security alternatives.
As further work, the method can be applied to different satellite channels and different power constraints for which average performance can be obtained in the case of considering short term stochastic fading at different frequencies. Losses due to different atmospheric phenomena can be easily incorporated to fully characterise the secrecy and communication performance and corresponding tradeoff for the given channel code under consideration.

\section*{Acknowledgment}
The second author was supported in part by JSPS Grant-in-Aid for Scientific Research (B) No.16KT0017 and for Scientific Research (A) No.17H01280,
the Okawa Research Grant and Kayamori Foundation of Informational Science Advancement.

\appendices

\section{Wiretap code construction}
\subsection{General construction}
Now, we discuss how to make a wiretap channel code to realize the security.
First, we prepare an auxiliary random variable $L_n$ subject to the uniform distribution on another set ${\cal L}_n$.
Then, we prepare an error correcting code.
That is, we prepare an encoder as a map 
$\phi_{e,n}$ from 
the set 
${\cal M}_n \times {\cal L}_n$ to the input alphabet ${\cal X}^n$,
and a decoder as a map $\phi_{d,n}$ from 
the input alphabet ${\cal X}^n$ to the set ${\cal M}_n \times {\cal L}_n$.

We also prepare 
another map $f_n $ from
${\cal M}_n \times {\cal L}_n$ to ${\cal M}_n$
satisfying the condition
\begin{align}
|f_n^{-1}(m)|=| {\cal L}_n|
\end{align}
for $m \in {\cal M}_n$. This map is often called a hash function.
The encoder of our wiretap code is given as follows.
When Alice intends to send a message $m$, 
using the auxiliary random variable $L_n$,
she generates a random variable $L_n'$ on $f_n^{-1}(m)$
because the cardinality of $f_n^{-1}(m)$ is the same as that of ${\cal L}_n$.
Then, she sends the alphabet $ \phi_{e,n} (L_n')$.
The decoder is given as the map $f_n \circ \phi_{d,n}$.

When ${\cal X}$ has a modular structure and 
the channels $W_{Y|X}$ and $W_{Z|X}$ satisfy covariant properties,
the channel is called symmetric.
Now, we consider the special symmetric case when 
${\cal X}=\{0,1\}=\bF_2$.

Now, we employ an algebraic error correcting code for the case with $ {\cal X}=\bF_2 $.
That is,
the map $\phi_{e,n}$ is given as a homomorphism from 
the set 
${\cal M}_n \times {\cal L}_n$ to the input alphabet ${\cal X}^n=\bF_2^n$,
where ${\cal M}_n$ and $ {\cal L}_n$ are given as $\bF_2^{k_{n}}$ and $\bF_2^{k_{n}'}$,
and 
the map $ f_n$ from
${\cal M}_n \times {\cal L}_n$ to ${\cal M}_n$ is given as an isomorphism.

Such an algebraic code is given as a pair of two sublinear spaces $ C_2\subset C_1 \subset \bF_2^n
$, where $C_1$ and $C_2$ are isomorphic to $\bF_2^{k_{n}+k_{n}'}$ and $\bF_2^{k_{n}'}$, respectively.
So, the map $\phi_{e,n}$ is given as the isomorphism from $\bF_2^{k_{n}+k_{n}'}$ to $C_1$.
Using the isomorphic from the quotient space $C_1/C_2 $ to $\bF_2^{k_{n}}$,
we define the map $ f_n$ from ${\cal M}_n \times {\cal L}_n$ to ${\cal M}_n$.
This type of encoder is called coset encoding.

\subsection{Randomized construction}
\label{sec:chann-resolv-with}

%Next, we consider randomized construction of coset coding.
%To this end, we denote by $q_{Z^n}$, the distribution induced at the output of the eavesdropper's channel when the input is an i.i.d. process of uniformly distributed bits $X$. 
In this section, we fix an algebraic error correcting code $(\phi_{e,n},\phi_{d,n})$.
Here, we choose integers $k_n$ and $k_n'$ such that
$k_n+k_n' $ is the message length of the given algebraic error correcting code $(\phi_{e,n},\phi_{d,n})$ and 
$k_n$ is the message length of our secrecy transmission.
So, $k_n'$ can be regarded as the sacrifice bit-length in our protocol.
Then, we consider a randomized hash function $F:\bF_2^{k_n+k_n'}\rightarrow\bF_2^{k_n}$
satisfying the following property:
\begin{align}
  \label{eq:condition_masahito}
\forall m\neq \forall m' \in\bF_2^{k_n+k_n'}
\quad
\P{F(m)=F(m')}\leq\frac{1}{2^{k_n}}.
\end{align}
This condition is called universal$_2$ \cite{Wegman1981}, and will be supposed in the remaining part.
%A typical universal2 hash function is given by Toeplitz matrix as follows.
A modified form of the Toeplitz matrices is also shown to be universal$_2$, 
which is given by a concatenation $(T(S), I)$ of the 
$k_n' \times k_n$ Toeplitz matrix $T(S)$ and the $k_n \times k_n$ identity matrix $I$ \cite{Hayashi2011},
where $S$ is the random seed to decide the Toeplitz matrix and belongs to $\bF_2^{k_n+k_n'-1}$.
The (modified) Toeplitz matrices are particularly useful in practice, because there exists an efficient multiplication algorithm using the fast Fourier transform algorithm with complexity $O(n\log n)$. 

When the random seed $S$ is fixed, 
the encoder for our wiretap code is given as follows.
By using the auxiliary random variable $L_n \in \bF_2^{k_n'}$,
the wiretap encoder is given as $\phi_{e,n}
\Big( 
\Big(
\begin{array}{cc} 
I & - T(S) \\
0 & I
\end{array}
\Big)
\Big(
\begin{array}{c} 
M \\
L_n
\end{array}
\Big)
\Big)$
because 
$
(I,T(S))
\Big(
\begin{array}{cc} 
I & - T(S) \\
0 & I
\end{array}
\Big)
= (I, 0)$.
(Toeplitz matrix $T(S)$ can be constructed as a part of circulant matrix. 
For example, the reference \cite[Appendix C]{Hayashi-Tsurumaru}
gives a method to give a circulant matrix.).
More efficient construction for universal2 hash function is discussed in \cite{Hayashi-Tsurumaru}.

So, the wiretap decoder is given as $Y^n \mapsto (I,T(S)) \phi_{d,n}(Y^n)$.
When $\phi_{d,n}$ can be efficiently performed like a LDPC code or a Polar code,
That is, as long as the algebraic error correcting code $(\phi_{e,n},\phi_{d,n})$
can be efficiently performed our code can be efficiently performed.

\section{Secrecy Exponent Function}
\label{sec:sec-exp-funct}
%Now, we give more precise security evaluation via channel resolvability.
Using the channel to eavesdropper described by a transition matrix $W_{Z|X}$,
we define the function
\begin{align}
&\psi(s|W_{Z|X},q_X)\nonumber \\
:=& \log \sum_{x,z} q_X(x) W_{Z|X}(z|x)^{1+s}
(\sum_x q_X (x) W_{Z|X}(z|x))^{-s} .
\end{align}
This function satisfies 
\begin{align}
\frac{\psi(s|W_{Z|X},q_X)}{s}\to I(X;Z)
\end{align}
as $s \to 0$.

Firstly, for simplicity, we consider the case when the main channel $W_{Y|X}$ is noiseless
and the channel to the eaves dropper is given as the $n$-fold extension of
$W_{Z|X}$.
In this case, the error correcting encoder $\phi_{e,n} $ is the identity map and $k_n+k_n'=n$.
Then, we have the following theorem.
\begin{Theorem}\label{TH1}
The above given code satisfies 
\begin{align}
{\mathbb{E}}_F  \text{S}_{\text{strong}}(M;Z^n)
\le \frac{1}{s} 2^{-s k_n'}e^{n \psi(s|W_{Z|X},p_{X,U})}
\end{align}
for $s \in [0,1]$, where $p_{X,U}$ is the uniform distribution.
\end{Theorem}
This theorem can be shown as a special case of \cite[(12)]{Hayashi2011}.
So, when 
\begin{align}
\label{eq:secrecy_conditB}
\frac{\psi(s|W_{Z|X},p_{X,U})}{s} < \frac{k_n}{n} \log 2,
\end{align}
the average of the leaked information 
${\mathbb{E}}_F  \text{S}_{\text{strong}}(M_n;Z^n)$ goes to zero exponentially.

Now, we proceed to the general case, i.e., the case when
the main channel $W_{Y|X}$ is noisy.
So, 
the error correcting encoder $\phi_{e,n} $ is not the identity map, and
$k_n+k_n'$ is smaller than $n$.
To discuss the general case, we introduce other functions
\begin{align}
E_0(s|W_{Z|X},q_X):=& \log 
\sum_{z}
(\sum_x q_X(x) W_{Z|X}(z|x)^{\frac{1}{1-s}} )^{1-s} \\
E_{0,\max}(s|W_{Z|X}):=& \max_{q_X}E_0(s|W_{Z|X},q_X).\label{H41}
\end{align}
This function satisfies 
\begin{align}
\frac{E_0(s|W_{Z|X},q_X)}{s}\to I(X;Z)
\end{align}
as $s \to 0$.
Further, we have the following theorem \cite[Theorem 7]{Hayashi2010}. 
\begin{Theorem}\label{T2UB}
The above given code
satisfies 
\begin{align}
{\mathbb{E}}_F  \text{S}_{\text{strong}}(M;Z^n)
\le \frac{1}{s} 2^{-s k_n'}e^{n E_{0,\max}(s|W_{Z|X})}. \label{H12-13}
\end{align}
for $s \in [0,1]$.
\end{Theorem}

Additionally, 
when $W_{Z|X}$ is symmetric, the convexity of the map 
$q_X \mapsto e^{E_0(s|W_{Z|X},q_X)}$ \cite[Lemma 1]{Hayashi2011}
and the symmetry yield the following theorem.

\begin{Theorem}\label{T2U}
When $W_{Z|X}$ is symmetric,
$E_{0,\max}(s|W_{Z|X})$ equals $E_0(s|W_{Z|X},p_{X,U})$.
That is, the maximum is achieved by the uniform distribution $p_{X,U}$ for $X$.
\end{Theorem}

Hence, we have \cite[(21)]{Hayashi2011}. 
\begin{align}
{\mathbb{E}}_F  \text{S}_{\text{strong}}(M;Z^n)
\le \frac{1}{s} 2^{-s k_n'}e^{n E_{0}(s|W_{Z|X},p_{X,U})}. \label{H12-13B}
\end{align}
for $s \in [0,1]$.

So, when $\frac{E_{0,\max}(s|W_{Z|X})}{s}$ is smaller than $(\frac{k_n'}{n})\log 2$,
the average of the leaked information 
${\mathbb{E}}_F  \text{S}_{\text{strong}}(M;Z^n)$ goes to zero exponentially.
Since this method can be applied to any algebraic error correcting code,
our method can be applied to the case when the code $(\phi_{e,n},\phi_{d,n})$ is a LDPC code or a Polar code.
So, our method provides an efficient wiretap code based on LDPC codes and Polar codes,
which attains the wiretap capacity for symmetric channel
because the maximum in \eqref{H41} is attained when $q_X$ is the uniform distribution.

Further, even when the auxiliary random number $L_n$ is not uniform,
a similar evaluation is available by slightly different coding given in \cite[Section XI]{Hayashi2012}.
Then, the above security evaluation holds with the replacement of $2^{-s k_n'} $ 
by $e^{s H_{1+s}(L_n)}$,
where $H_{1+s}(X)$ is the R\'{e}nyi entropy defined as
\begin{align}
H_{1+s}(X):=-\frac{1}{s}\log\sum_{x}P_X(x)^{1+s}  .
\end{align}
These evaluations still hold even when the output system of $W_{Z|X}$ is continuous,
e.g., the AWGN channel \cite[Appendix D]{Hayashi2012}.
In particular, the AWGN channel with binary input can be regarded as a symmetric channel.

\section{Secrecy Capacity of RF Gaussian Channel}
In this section, using a certain regularized parameter $\gamma_{g,0}$ defined in (16) of the main body,
we derive the secrecy capacity with the BPSK scheme
when Bob's and Eve's observed signals are given as
%???Dividing by the common path loss channel coefficient in Bob's and Eve's deterministic path losses, we can re-write the model so that Bob's signal is the reference for Eve's signal
\begin{equation}\label{eq:Gaussig2}
\begin{split}
Y&=X+\sqrt{n_{B}}X_1;\\
Z&=\gamma_{g,0} X+\sqrt{n_{B}}X_2;
\end{split}
\end{equation}
with $X_1$ and $X_2$, zero-mean circular complex Gaussian random variables with unit variance,
$n_B$ is the noise power level at Bob's receiver.
The independent Eve's and Bob's channels are discrete input and continuous AWGN output channels. The input random variable $X$ takes $2^{m}$-PSK modulation vales as
${\cal X}:=\left\{\exp\left[{j\frac{2\pi\left(i-1\right)}{2^{m}}}\right],i=1,\ldots,2^{m}\right\} $.

The mutual information for Bob's channel is 
\begin{align*}
I(X;Y)=\int_{-\infty}^{\infty}\sum_{x\in{\cal X}}W_{Y|X}\left(y|x\right)q_{X}\left(x\right)\log_{2}\frac{W_{Y|X}\left(y|x\right)}{W_{Y}\left(y\right)}dy
\end{align*}
where 
\begin{align*}
W_{Y}\left(y\right):=\sum_{x}W_{Y|X}\left(y|x\right)q_{X}\left(x\right).
\end{align*}
The mutual information for Eve's channel is 
\begin{align*}
I(X;Z)=\int_{-\infty}^{\infty}\sum_{x\in{\cal X}}W_{Z|X}(z|x)q_{X}(x)\log_{2}\frac{W_{Z|X}(z|x)}{W_{Z}(z)}dz
\end{align*}
where 
\begin{align*}
W_{Z}(z):=\sum_{x}W_{Z|X}(z|x)q_{X}(x).
\end{align*}

Since the channels 
$W_{Z|X}$ and $W_{Y|X}$ are Gaussian channels and their difference is only the power of their noise,
secure communication is possible if and only if 
the channel $W_{Z|X}$ is given as a degraded channel of $W_{Y|X}$,
i.e., 
there exists a channel $W_{Z|Y}$ such that
\begin{align}
W_{Z|X}(z|x)=\int_{-\infty}^{\infty} W_{Z|Y}(z|y) W_{Y|X}(y|x) dy.
\end{align}
This condition is equivalent to the 
the following \textbf{spatial stochastic degradation condition}
for the constant coefficient $\gamma_{g,0}$;
%introduced in Section III.B of the main body;
\begin{equation}
\gamma_{g,0} < 1,
\label{eq:secrecy_condit}
\end{equation}
which is addressed in Section III-B of the main body.

\if0
In this case, the channel $W_{Z|Y}$ is given as
\begin{align}
W_{Z|Y}(z|y):= 
\frac{1}{\sqrt{2\pi n_{B}(-\gamma_{g,0}^2)}}
e^{-\frac{\left(z-\gamma_{g} y \right)^{2}}
{2 n_{B}(-\gamma_{g,0}^2)}},
\end{align}
which means that $Z=\gamma_{g,0} Y+\sqrt{n_B(-\gamma_{g,0}^2)}X_3$ with 
a zero-mean circular complex Gaussian random variable $X_3$ with unit variance that is independent of $X_1$.
This characterization can be shown as follows.
Using the first equation of \eqref{eq:Gaussig2}, we have
\begin{align*}
Z&=\gamma_{g,0} Y+\sqrt{n_B(-\gamma_{g,0}^2)}X_3\\
&=\gamma_{g,0} X+ \gamma_{g,0} \sqrt{n_{B}}X_1+\sqrt{n_B(-\gamma_{g,0}^2)}X_3.
\end{align*}
Letting $X_2$ be $\frac{1}{ n_B} (\gamma_{g,0} \sqrt{n_{B}}X_1+\sqrt{n_B(-\gamma_{g,0}^2)}X_3)$, which is 
another zero-mean circular complex Gaussian random variable with unit variance.
So, we obtain the second equation of \eqref{eq:Gaussig2}.
\fi

Using this property, we can simplify the capacity formula
\begin{align}
  \label{eq:secrecy_capacity}
  C_s = \max_{p_{VX}}\left(I(V;Y)-I(V;Z)\right)_+.
\end{align}
In the formula \eqref{eq:secrecy_capacity},
we have the Markov chain $V-X-Y-Z$, which implies that
\begin{align}
&I(V;Y)-I(V;Z)= I(V;YZ)-I(V;Z) \nonumber \\
=& I(V;Y|Z)
\le I(X;Y|Z)
= I(X;Y)-I(X;Z).
\end{align}
Here, the conditional mutual information
$I(X;Y|Z)$ is given as
\begin{align*}
I(X;Y|Z) %\nonumber \\
=&\int_{-\infty}^{\infty}\sum_{x\in{\cal X}}
W_{Z|Y}(z|y) W_{Y|X}(y|x)q_{X}(x) \nonumber \\
& \cdot \log_{2} \frac{W_{Y|Z X}(y|z,x)}{W_{Y|Z}(y|z)}dz dy ,
\end{align*}
where
\begin{align*}
W_{Y|Z X}(y|z,x)&:=
W_{Z|Y}(z|y) W_{Y|Z}(y|z)/W_{Z|x}(z|x) \\
W_{Y|Z}(y|z)&:=
W_{Z|Y}(z|y) W_{Y}(y)/W_Z(z).
\end{align*}

Hence, the capacity given in \eqref{eq:secrecy_capacity} is simplified as
\begin{align}
C_s 
= \max_{q_{X}}\left(I(X;Y)-I(X;Z)\right)_+
= \max_{q_{X}}\left(I(X;Y|Z)\right)_+.
\label{eq:Cs_max}
\end{align}

%\begin{figure}[tbh]
%\centering\includegraphics[scale=0.5]{CsBIAWGN}
%\protect\caption{Secrecy capacity given by (\ref{eq:Cs_BI-AWGN}). It is represented as a function of $\gamma_{g,0}$ and $E_0/N_0 = E_0/n_B$ with $E_0=1$.}\label{CsBIAWGN}
%\end{figure}

To express the dependence with respect to $q_X$,
we denote the conditional mutual information by $I(X;Y|Z)_{q_X}$.
Notice that the mutual information $I(X;Y|Z)_{q_X}$ is concave
for $q_X$.
Let $g$ be the rotation in ${\cal X}$ by the multiplication of 
$\exp\left[{j\frac{2\pi}{2^{m}}}\right]$.
For a distribution $q_X$, we define the distribution $g(q_X)$ as 
$g(q_X)(x):=q_X(g(x))$.
Since $I(X;Y|Z)_{g(q_X)}=I(X;Y|Z)_{q_X} $, the mutual information $I(X;Y|Z)_{q_X}$ is upper bounded by the mutual information $I(X;Y|Z)_{p_X,U}$ with the uniform distribution $p_{X,U}$ as
\begin{align}
&I(X;Y|Z)_{q_X}
= \sum_{i=0}^{2^m-1}2^{-m} I(X;Y|Z)_{g^{i}(q_X)} \nonumber \\
\le &  I(X;Y|Z)_{\sum_{i=0}^{2^m-1}2^{-m} g^{i}(q_X)}
=I(X;Y|Z)_{p_{X,U}}.
\end{align}
The final equation follows from the fact that
the distribution $\sum_{i=0}^{2^m-1}2^{-m} g^{i}(q_X)$
is the cyclic mixture of $q_X$ on ${\cal X}$, which equals the uniform distribution 
$p_{X,U}$ on ${\cal X}$.
In this case, the maximization in \eqref{eq:Cs_max} is achieved by
the uniform distribution due to the symmetry. 
That is,
\begin{align}
C_s =I(X;Y|Z)_{p_{X,U}}.
\label{eq:Cs_expression}
\end{align}

In general, the bit error probability of $2^{m}$-PSK is difficult
to obtain for an arbitrary integer $m$. We assume from now $m=1$
so that $X$ takes values in $\{-1,1\}$, which is called BPSK modulation. 
Denoting as $h(X)$ the differential entropy of a random variable $X$, 
the secrecy capacity for BPSK input can be computed as the function of 
$\gamma_{g,0}$ and $n_B$
in the following way;
\begin{align}
&C_s(\gamma_{g,0} , n_B) \nonumber\\
=&I(X;Y)-I(X;Z) \nonumber\\
=&h(X) - h(X|Y) -h(X) + h(X|Z)\nonumber\\
=&h(X|Z) -  h(X|Y) \nonumber\\
=&h(X) - h(Z) + h(Z|X) - [ h(X) - h(Y) + h(Y|X) ]\nonumber\\
=&-h(Y|X) + h(Y) - [ - h(Z|X) + h(Z) ]\nonumber\\
=&\int_{-\infty}^\infty \frac{1/2}{\sqrt{2\pi n_B}}u\Big[e^\frac{-(y+1)^2}{n_B} + e^\frac{-(y-1)^2}{n_B} \Big]dy + \frac{1}{2}\log(2 \pi e n_B)\nonumber\\
&- \int_{-\infty}^\infty \frac{1/2}{\sqrt{2\pi  n_B}}
u\Big[e^\frac{-(z+\gamma_{g,0})^2}{ n_B} 
+ e^\frac{-(z-\gamma_{g,0})^2}{n_B} \Big] dz
%\nonumber\\ &
- \frac{1}{2}\log(2 \pi e n_B)\nonumber\\
=&\int_{-\infty}^\infty  \frac{1}{\sqrt{8\pi n_B}}u\Big[e^\frac{-(y+1)^2}{n_B} 
+ e^\frac{-(y-1)^2}{n_B} \Big]dy\nonumber\\
&-\int_{-\infty}^\infty  \frac{1}{\sqrt{8\pi n_B}}u\Big[e^\frac{-(z+\gamma_{g,0})^2}{n_B} 
+ e^\frac{-(z-\gamma_{g,0})^2}{n_B} \Big]dz.
\label{H42}
\end{align}
where $u(x):= - x \log x$. Therefore, we obtain the formula (18) of the main body for the secrecy capacity.

%\begin{figure}[tbh]
%\centering
%\includegraphics[scale=0.4]{ConditionRF}
%\caption{Visualization of positive capacity when degradation condition \eqref{eq:secrecy_condit} for the RF channel holds in the case of soft decoding.}
%\label{ConditionRF}
%\end{figure}

%For positive secrecy condition is \eqref{eq:secrecy_condit} as shown in Fig. \ref{ConditionRF}. 
This secrecy capacity is asymptotically achievable by the wiretap code given in Section \ref{sec:chann-resolv-with}. 
The \textbf{infinite-length condition of reliability} holds since our proposed coding can be applied to any algebraic error correcting code. Therefore, the code $(\phi_{e,n},\phi_{d,n})$ can be a LDPC code or a Polar code, 
which are capacity achieving for this symmetric channel and hence asymptotically attain the wiretap secrecy capacity because the maximum in \eqref{eq:Cs_max} is attained when $q_X$ is the uniform distribution. The \textbf{infinite-length condition of secrecy} also holds as explained in Section \ref{sec:sec-exp-funct}, since condition \eqref{eq:secrecy_condit} is not constrained by the underlying modulation and detection methods, i.e. when we randomly choose $F$, the quantity $\text{S}_{\text{strong}}(M;Z^n)$ is sufficiently small with high probability.
\ifCLASSOPTIONcaptionsoff
  \newpage
\fi


%\end{IEEEbiography}


%\newpage




%\vfill


\end{document}